\documentclass[aps,twocolumn,a4paper,superscriptaddress]{revtex4}
\usepackage{tensor}
\usepackage{graphicx}
\usepackage{amsmath}
\usepackage{amssymb}
\usepackage{enumerate}
\usepackage{subfigure}
\usepackage{tabularx}
\usepackage[colorlinks=true, pdfstartview=FitV, linkcolor=blue, citecolor=red, urlcolor=black, breaklinks=true]{hyperref}
\usepackage{amsthm}
\usepackage{epstopdf}
\newcommand{\be}{\begin{equation}}
\newcommand{\ee}{\end{equation}}
\newcommand{\ben}{\begin{eqnarray}}
\newcommand{\een}{\end{eqnarray}}
\newcommand{\bes}{\begin{subequations}}
\newcommand{\ees}{\end{subequations}}
\def\bal#1\eal{\begin{align}#1\end{align}}

\newcommand{\sech}{{\rm sech}}
\newcommand{\LL}{{\mathcal L}}

\newcommand{\Lag}{\mathcal{L}} 
 
\newcommand{\pu}{\mathrm{\partial_{\mu}}}

\begin{document}
\title{Mechanism to induce geometric constriction on kinks and domain walls}
\author{A.J. Balseyro Sebastian}\affiliation{Departamento de F\'\i sica, Universidade Federal da Para\'\i ba, 58051-970 Jo\~ao Pessoa, PB, Brazil}
\affiliation{IUFFyM, Universidad de Salamanca, Spain}
\author{D. Bazeia}\affiliation{Departamento de F\'\i sica, Universidade Federal da Para\'\i ba, 58051-970 Jo\~ao Pessoa, PB, Brazil}
\author{M.A. Marques}\affiliation{Departamento de Biotecnologia, Universidade Federal da Para\'\i ba, 58051-900 Jo\~ao Pessoa, PB, Brazil}
\begin{abstract}
We investigate scalar field theories in the multifield scenario, focusing mainly on the possibility to smoothly build internal structure and asymmetry for kinks and domain walls. The procedure requires the inclusion of an extra field which is associated to a function that modifies the dynamics of the other fields. We investigate minimum energy configurations, which support first order equations compatible with the equations of motion. The extra field allows a transition which is guided by a parameter that connects the standard solution to another one, geometrically constrained, mimicking the effects of geometrical constrictions in magnetic materials. 
\end{abstract}
\maketitle


\section{Introduction}
Topological objects that arise in high energy physics under the presence of real scalar fields \cite{B1,B2,B3} are of current interest due to the several applications, for instance, to describe thick branes in distinct scenarios; see, e.g., Refs. \cite{BW00,BW1,BW2}, the review \cite{BW3} and other more recent works \cite{BW0,BW4,BW5,BW6,BW7} and references therein. In particular, in \cite{BW0} the investigation deals with self-gravitating kink-like solutions in a two-dimensional dilaton gravity, in \cite{BW4} the study describes a mechanism to control the internal structure of thick brane, in \cite{BW5} the authors deal with thick brane in Rastall gravity, and in \cite{BW6,BW7} the investigations concern aspects of quantum gravity and computation of holographic entanglement entropy in Karch-Randall brane \cite{KRB} in anti-de Sitter space.

These topological structures are also of interest in condensed matter, with the wide range of applications described, for instance, in \cite{KCM1,A1,A2,A3,A4} and in references therein. In \cite{KCM1} the authors show how the presence of geometric constriction in the
magnetic material is capable of controlling the way the
magnetization localizes inside the wall structure.
In \cite{A1} the investigation deals with the possibility to control the domain wall polarity of a magnetic domain wall by electric pulses, in \cite{A2} the study focuses on minimal strips and shows that when the strip is nonorientable, the system is mapped onto a scalar $\phi^4$ model on a line bundle over the circle, where the topological defect becomes a topologically protected domain wall. Moreover, in \cite{A3} the work investigates interlayer Dzyaloshinkii-Moriya interaction, which can break degeneracy between Bloch wall chiralities, also inducing asymmetries in the wall component of the Bloch domain walls in multilayers, with the asymmetry being a result of the combined effect of the demagnetization field and an interlayer Dzyaloshinkii-Moriya interaction, being strongly related to film thickness and structural ordering, and in \cite{A4} the authors suggest a mechanism to manipulate magnetic domain walls in ferrimagnetic or ferromagnetic multiferroics using electric fields. 

In materials science one can also discuss ferroelectric elements \cite{FE0}, where the negative capacitance may be connected to the double-well shape of the ferroelectric polarization, and multiferroic materials, in which coexistence of magnetic and ferroelectric order can provide an efficient route to control magnetism by electric fields \cite{MS1} and the exotic polarization profiles that can arise at domain walls with multiple order parameters and the different mechanisms that lead to domain-wall polarity in non-polar ferroelastic materials \cite{MS2}.

The main motivation of the present work is based on the experimental result uncovered in Ref. \cite{KCM1}, concerning the presence of geometric constrictions. We have studied this possibility theoretically in \cite{multikink}, finding an interesting connection between the geometric mechanism and the internal structure of the kinklike configuration. Here we want to further explore this possibility, adding new degrees of freedom and changing the way the geometric constriction acts in the system, modifying the mechanism in order to continuously control the geometric constriction to smoothly modify the internal structure and symmetry of the localized configuration. 

\section{Procedure}
We start the work introducing the Lagrangian density associated to the presence of two real scalar fields in $(1,1)$ spacetime dimensions. The system is described by natural units and dimensionless fields and coordinates, in the form
	\begin{equation}\label{lagr}
		\Lag = \frac{1}{2}f(\psi)\pu \phi \partial^{\mu} \phi + \frac{1}{2}\pu \psi \partial^{\mu} \psi  - V(\phi,\psi),
	\end{equation}
where $\phi$ and $\psi$ are real scalar fields coupled through the potential $V(\phi,\psi)$ and the non negative function $f(\psi)$. The model was previously investigated in Ref.~\cite{multikink}, where the function $f(\psi)$ was shown to engender the ability to simulate the presence of geometrical constrictions in kinklike structures, in a way similar to the experimental investigation described in \cite{KCM1}, which unveiled the direct influence of geometric constriction on the profile of the magnetization in the magnetic material. Moreover, inspired by the experimental investigation \cite{A1} on the possibility to control the domain wall polarity in a magnetic material in the presence of electric pulses, in \cite{BMM} one added fermions to the above model \eqref{lagr} to study how the presence of geometric constrictions may influence the behavior of fermions in the model. 

The model \eqref{lagr} can be considered to study minimum energy solutions. For static fields, we follow Ref.~\cite{bogo} and use results of \cite{multikink} to write first order differential equations 
\begin{equation}\label{fo1}
			\frac{d\phi}{dx}=\pm \frac{W_{\phi}(\phi, \psi)}{f(\psi)},\qquad
			\frac{d\psi}{dx}=\pm W_{\psi}(\phi,\psi)
	\end{equation}
that solve the equations of motion when the potential has the form
\begin{equation}\label{potW}
		V(\phi,\psi)=\frac12\frac{W_{\phi}^2}{f(\psi)} + \frac12\,{W_{\psi}^2},
	\end{equation}
where $W=W(\phi,\psi)$ is an auxiliary function and the subscripts denote partial derivatives, that is, $W_\phi=\partial W/\partial \phi$ and $W_\psi=\partial W/\partial \psi$. Notice that, in general, the above equations are coupled, and $W_\phi$ and $W_\psi$ may depend on both fields. Moreover, as one knows the energy of the field configurations that solve the first order equations is minimized to $E = \left|W(v_+,w_+)-W(v_-,w_-)\right|$, where $v_\pm = \phi(\pm\infty)$ and $w_\pm=\psi(\pm\infty)$. The energy density can be written as the sum of two contributions, $\rho = \rho_1 + \rho_2$, where
\be
\rho_1 = \frac{W_\phi^2}{f(\psi)},\qquad \rho_2 = W_\psi^2.
\ee
As it was shown in Ref.~\cite{multikink}, the model becomes particularly interesting in the case $W(\phi,\psi)=g(\phi)+ h(\psi)$; this happens because the field $\psi$ can be treated independently, and can be used to feed $f(\psi)$ in the first order equation for $\phi$. Indeed, we can rewrite Eqs.~\eqref{fo1} as
\begin{equation}\label{fo}
			\frac{d\phi}{dx}=\pm \frac{g_{\phi}(\phi)}{f(\psi(x))},\qquad \frac{d\psi}{dx}=\pm h_\psi(\psi).
	\end{equation}
Moreover, it is possible to introduce a geometrical coordinate, $\xi$, as
\be\label{xi}
\xi = \int\frac{dx}{f(\psi(x))},
\ee
so the first order equations now read
\be\label{foxi}
\frac{d\phi}{d\xi}=\pm g_{\phi}(\phi),\qquad \frac{d\psi}{dx}=\pm h_\psi(\psi).
\ee
This means that $\psi$, which is independent of $\phi$, can be used to simulate geometrical constriction on $\phi$, mimicking the effects that arise in the magnetic material investigated in Ref.~\cite{KCM1}. 

\section{Two fields} The above general results have been described previously in Ref.~\cite{multikink}, and in the present study we want to disclose other important new features, which were never seen before. The main motivation is to construct localized configurations with internal structure, which can also be symmetric or asymmetric, of current interest to high energy physics and with potential application to condensed matter at the nanometric scale. To develop our investigation, we first take
\be\label{hpsi}
h(\psi) = \alpha\psi-\frac\alpha3\psi^3,
\ee
with $\alpha$ being a positive real parameter. This leads us to the $\psi^4$ model which is a prototype of the Higgs field, of current interest in high energy physics \cite{B1}; it can also be used to represent the Ginzburg-Landau model of second order phase transition, with engenders a diversity of applications in nonlinear science, in particular, in ferroelectric material, to unveil the double-well energy landscape in a ferroelectric layer \cite{FE} and also in ferromagnetic materials, as in \cite{FM}, in which constriction in a trilayer system has been used to create skyrmions at room temperature. 

The choice in \eqref{hpsi} leads us to get
\be\label{solpsi}
\psi = \tanh(\alpha (x-x_0)),\qquad\rho_2 = \alpha^2\sech^4(\alpha (x-x_0)).
\ee
In the above expressions, $x_0$ is an integration constant that determines the point in which $\psi$ vanishes. Due to translational invariance, this is not important, so we fix $x_0$ at our convenience. Moreover, after integrating the energy density $\rho_2(x)$, one gets $E_2 =4\alpha/3$.

Following the main motivation of the present work, the above solution has to be used to feed the function $f(\psi(x))$. It is worth commenting here that, the geometrical coordinate in Eq.~\eqref{xi} depends only on $f(\psi(x))$, so we can choose  specific functions in order to get the desired profile of the other field configuration. This means that $f(\psi)$ will ultimately be a function of the spatial coordinate $x$, through the field $\psi$, but this interesting procedure is achieved without breaking translational invariance of the system. We then go on to investigate the first order equation for $\phi$, taking the function $g(\phi)$ in the $\phi^4$ form, as $g(\phi) = \phi-\phi^3/3.$
This leads us to
\be\label{solphi}
\phi = \tanh(\xi), \qquad \rho_1 = \frac{\sech^4(\xi)}{f(\tanh(\alpha (x-x_0))}.
\ee
These results are now governed by the new geometric coordinate $\xi$, given by Eq.~\eqref{xi} and so explicitly dependent on the specific form of $f(\psi)$. We notice that although one has already chosen both $g(\phi)$ and $h(\psi)$, there is still room to choose $f(\psi)$, to describe systems of current interest. This is our main concern from now on, where we deal with functions that do not modify the tail of the solution, allowing us to integrate the above energy density to get $E_1=4/3$. Since this is the energy of the standard $\phi^4$ model, our proposal is then to modify the profile of the solution without changing its total energy.  

Let us now introduce a model which is capable of smoothly changing the internal structure and the symmetry of the $\phi$ field. We consider the function
\be\label{flambda1}
f(\psi) = \frac{1+\lambda}{1+\lambda\psi^2},
\ee
where $\lambda$ is a nonnegative real parameter. This choice shows that as $x\to\pm\infty$, one gets $\psi\to\pm1$ and then $f$ becomes unity: asymptotically, then, the field $\phi$ behaves as in the standard $\phi^4$ model. On the other hand, in the center of the solution, for $x\to x_0$, $\psi\to0$ and we have $f=1+\lambda$. This means that the above function smoothly modifies the core of $\phi$, keeping its tails unchanged. Also, for the above $f(\psi)$, the geometrical coordinate in Eq.~\eqref{xi} becomes
\be\label{xi1}
\xi = \xi_0 + x-x_0 - \frac{\lambda}{\alpha(1+\lambda)}\tanh(\alpha (x-x_0)),
\ee
where $\xi_0$ is an integration constant. Notice that for $\lambda=0$, one gets $\xi=x-x_0+\xi_0$, which recovers the standard solution. On the other hand, for $\lambda\to\infty$, one gets $\xi = x-x_0 +\xi_0-\tanh(\alpha(x-x_0))/\alpha$, which induces null derivative in the solution, as previously described in Ref.~\cite{multikink}. We can expand $\xi$ near the point $x=x_0$ to get
\be
\xi=\xi_0 + \frac{1}{1+\lambda}(x-x_0) + \frac{\alpha^2\lambda}{3(1+\lambda)}(x-x_0)^3 + \mathcal{O}\left[(x-x_0)^5\right].
\ee
This shows that, as the parameter $\lambda$ increases, the linear term in $x-x_0$ becomes smaller and smaller, tending to zero in the limit $\lambda\to\infty$. We see that as $\lambda$ increases, the geometric constriction which is absent at $\lambda=0$, begins to appear and becomes fully active at very large values. In this sense, $\lambda$ controls the importance of the geometric constriction on the localized structure, which is of direct use to applications of practical interest. It can be considered, for instance, to continuously manipulate the behavior of the  magnetization in magnetic materials, in a way similar to the previously unveiled in Ref. \cite{KCM1}.   
		\begin{figure}[t!]
		\centering
		\includegraphics[width=4.2cm]{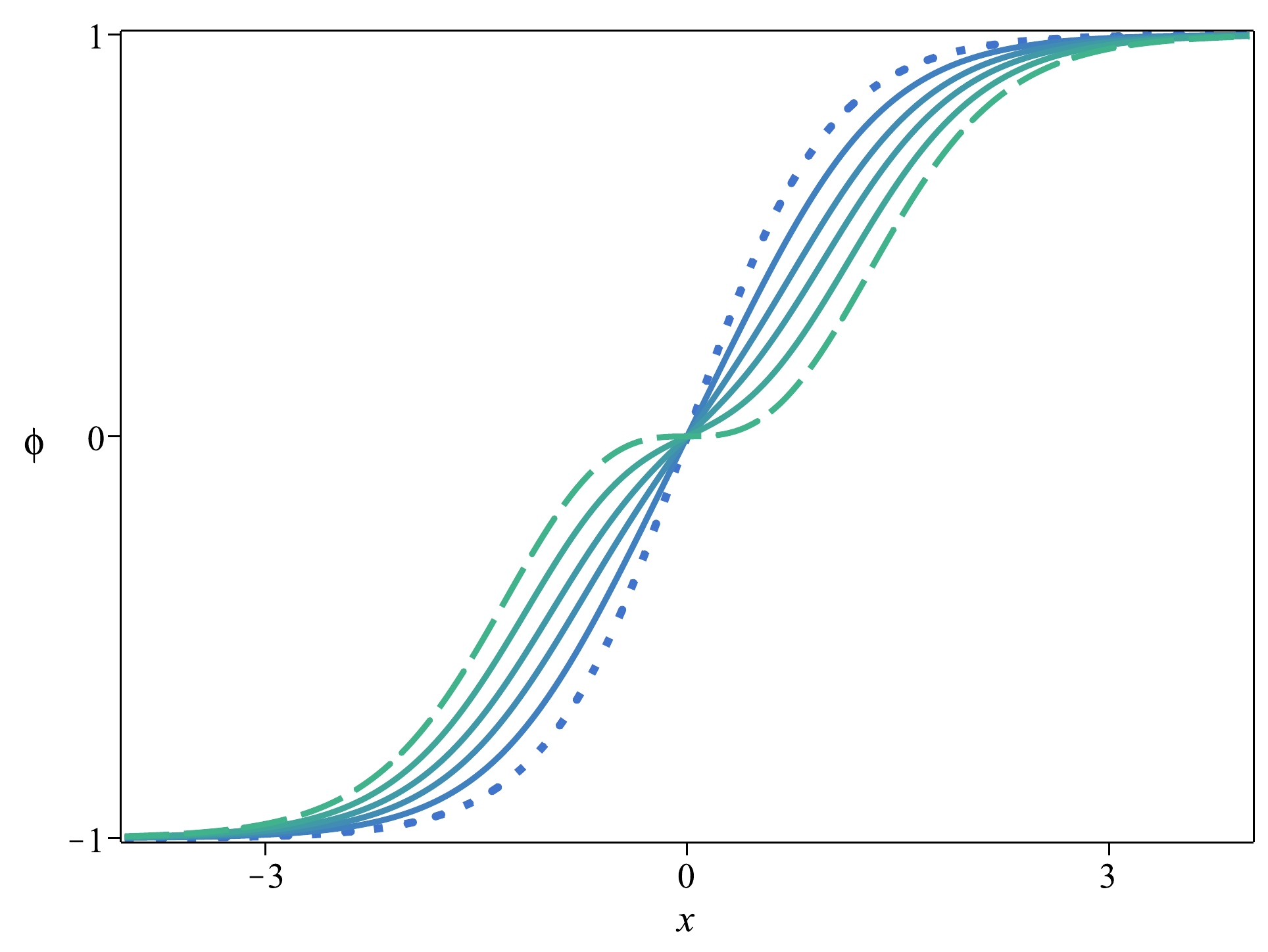}
		\includegraphics[width=4.2cm]{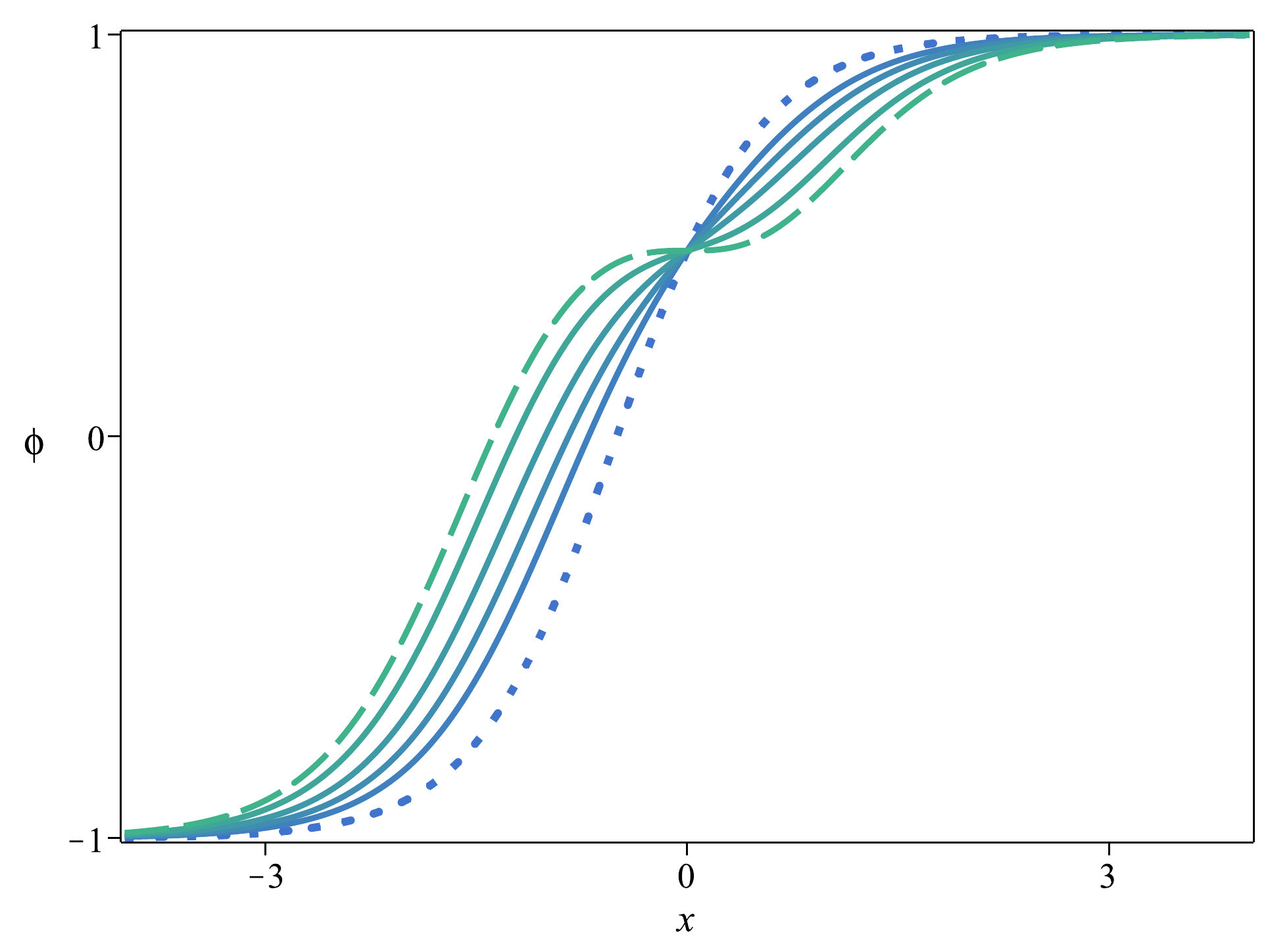}
		\includegraphics[width=4.2cm]{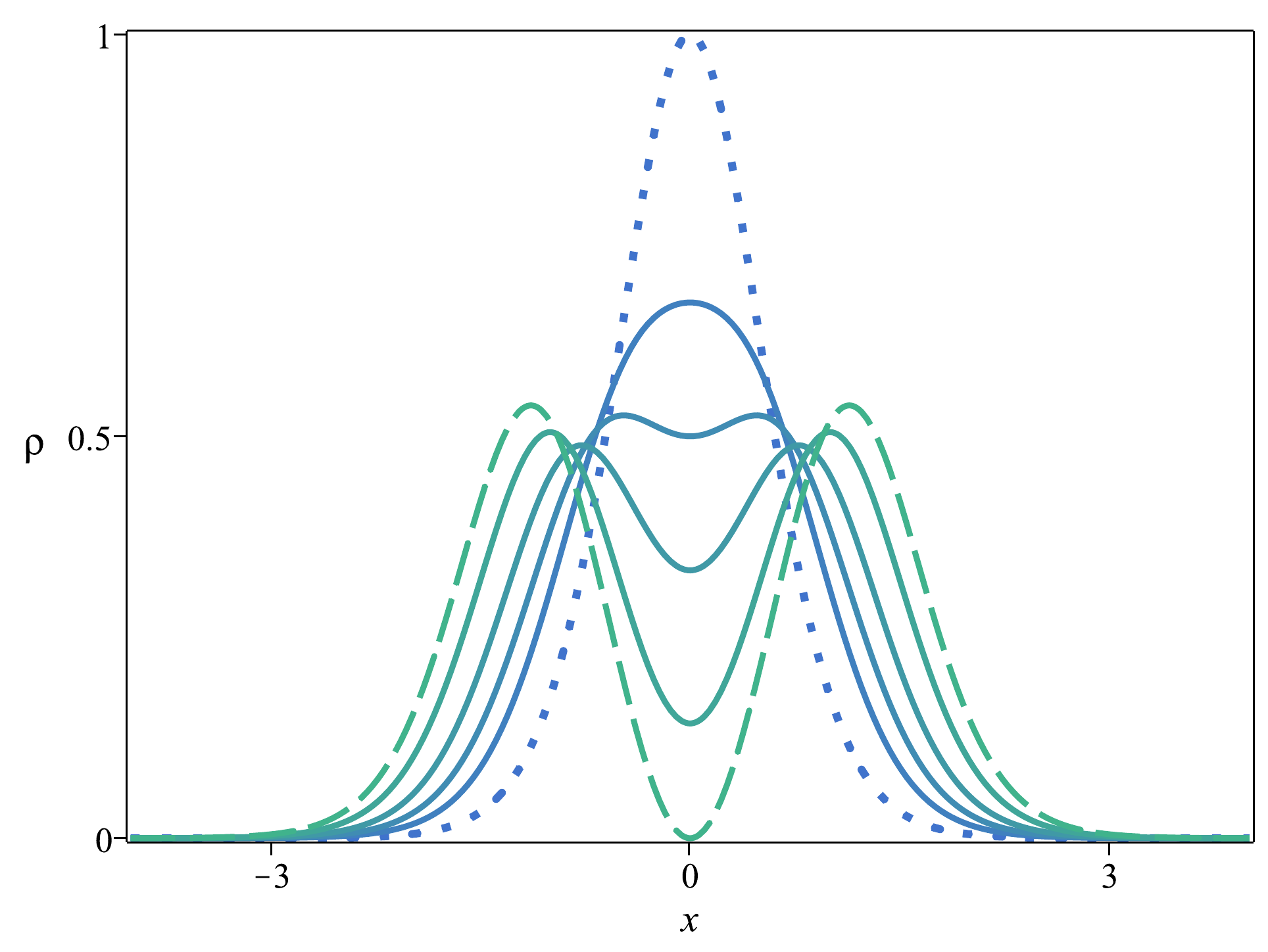}
		\includegraphics[width=4.2cm]{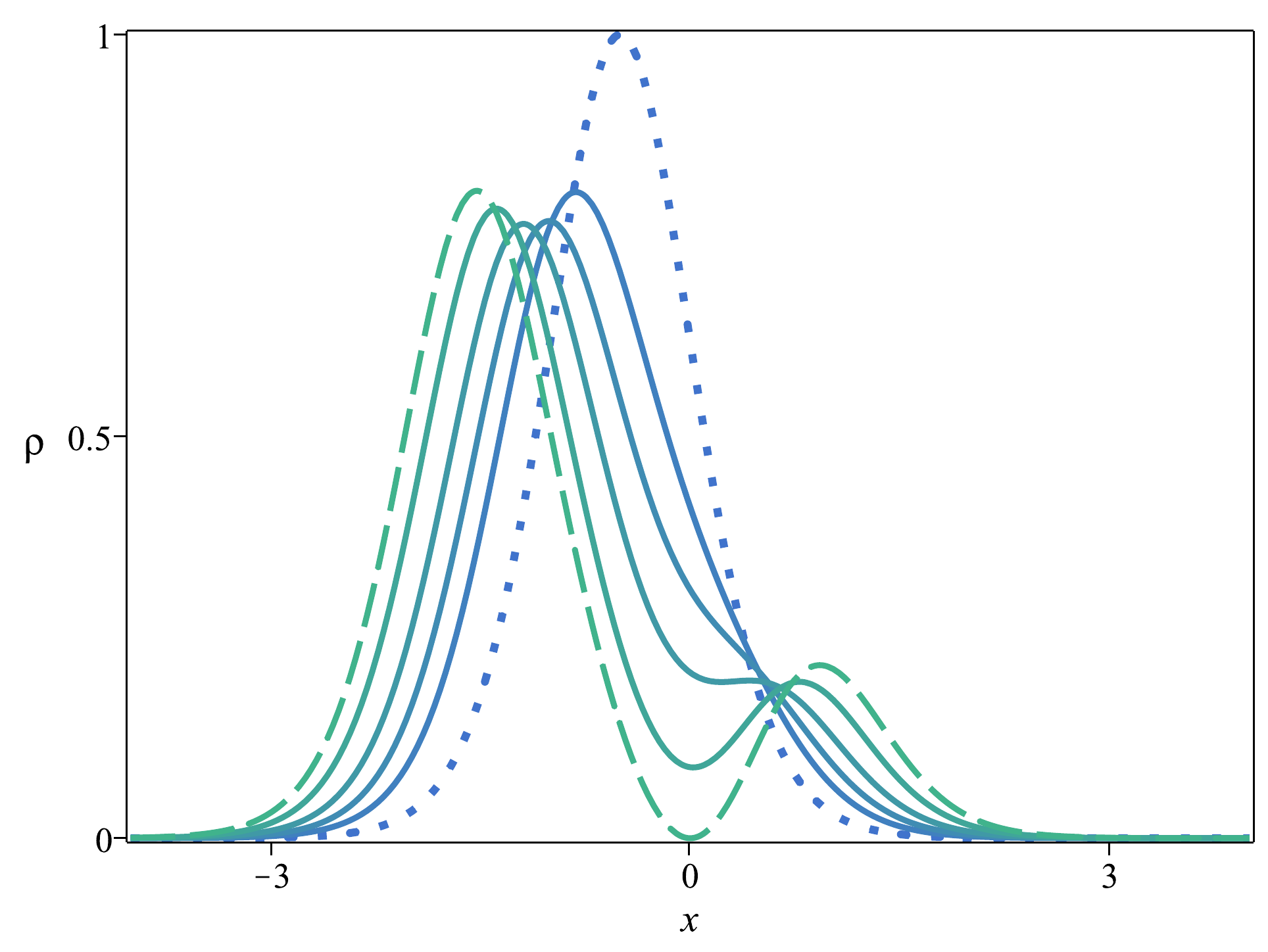}
		\caption{The solution $\phi$ (top) and the energy density $\rho_1$ (bottom) in Eq.~\eqref{solphi} for the geometrical coordinate in Eq.~\eqref{xi1} for several values of $\lambda$, with $\alpha=1$, $x_0=0$ and $\xi_0=0$ (left) and $1/2$ (right). The dotted lines represent the case with $\lambda=0$ and the dashed ones, the limit $\lambda\to\infty$.}
		\label{figsolrho1}
		\end{figure}

Since $x_0$ only shifts the solutions in the $x$ axis, we take $x_0=0$ and plot, in Fig.~\ref{figsolrho1}, solution for $\phi$ and its respective energy density $\rho_1$ in Eq.~\eqref{solphi}. There are two features which we want to highlight here. First, we see that an inflection point with null derivative tends to appear in the solution as $\lambda$ increases. Regarding the energy density, in the case $\xi_0=0$ we see that the central maximum in the energy density tends to become a minimum as $\lambda$ gets larger and larger, showing the action of the function \eqref{flambda1} in the geometry of the localized structure defined by $\phi$. The central point is a  maximum for $\lambda\leq\lambda_*$ or a minimum for $\lambda>\lambda_*$, with
$\lambda_*=\beta^{1/3}/(3\alpha)+\alpha/(3\beta^{1/3}) -2/3,$
and $
\beta=\alpha\!\left(\alpha^2+27+3\sqrt{3(2\alpha^2+27)}\right).$
The second feature we stress here is the presence of an asymmetry in the solution and energy density in the case of non-vanishing $\xi_0$, for any $\lambda>0$. The presence of $\xi_0$ shifts the inflection point upwards or downwards the origin in the solution and leads to an energy density formed by two lumps of different heights. The localized structure is then asymmetric. 

\begin{figure}[t!]
		\centering
		\includegraphics[width=4.2cm]{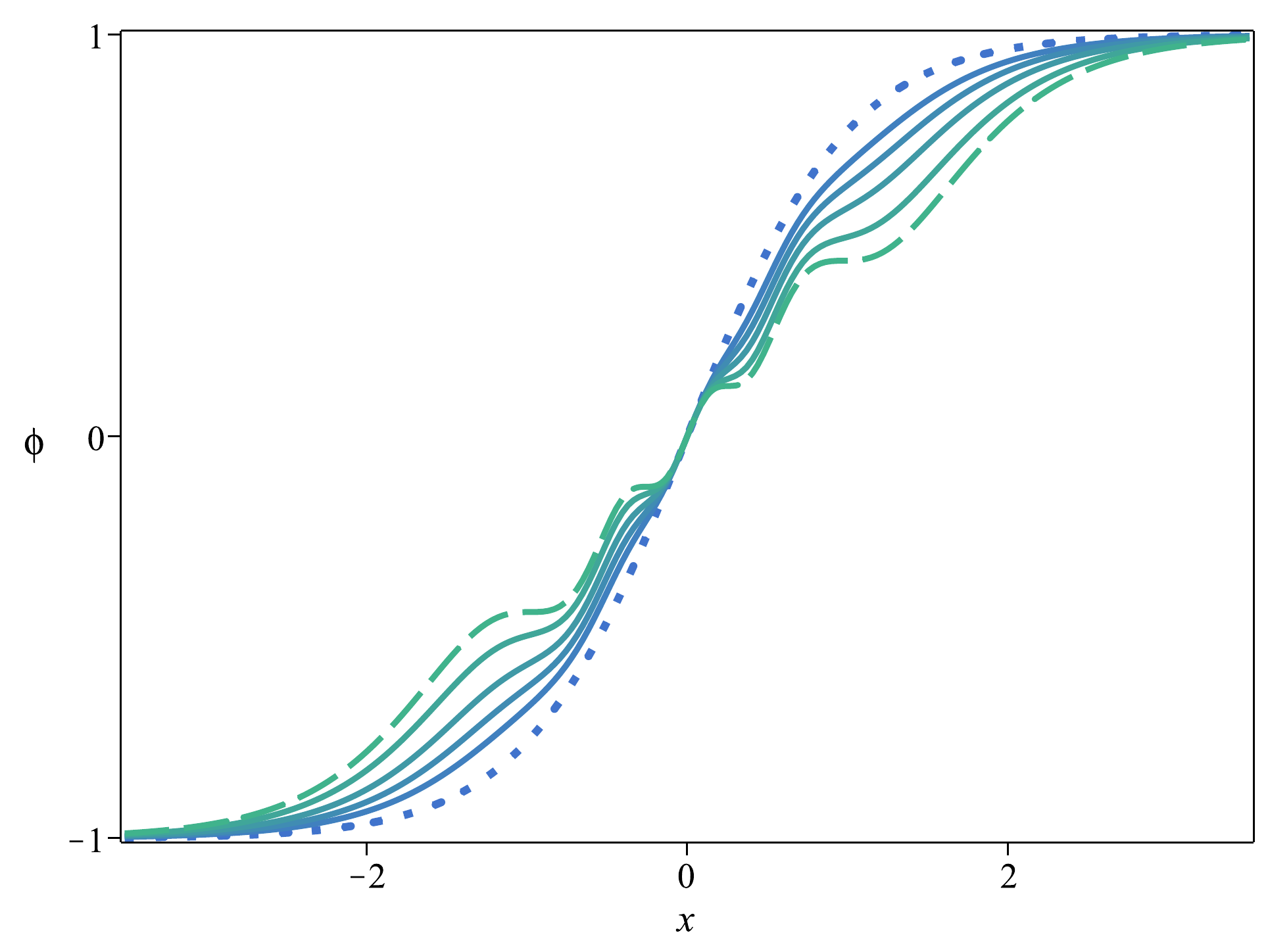}
		\includegraphics[width=4.2cm]{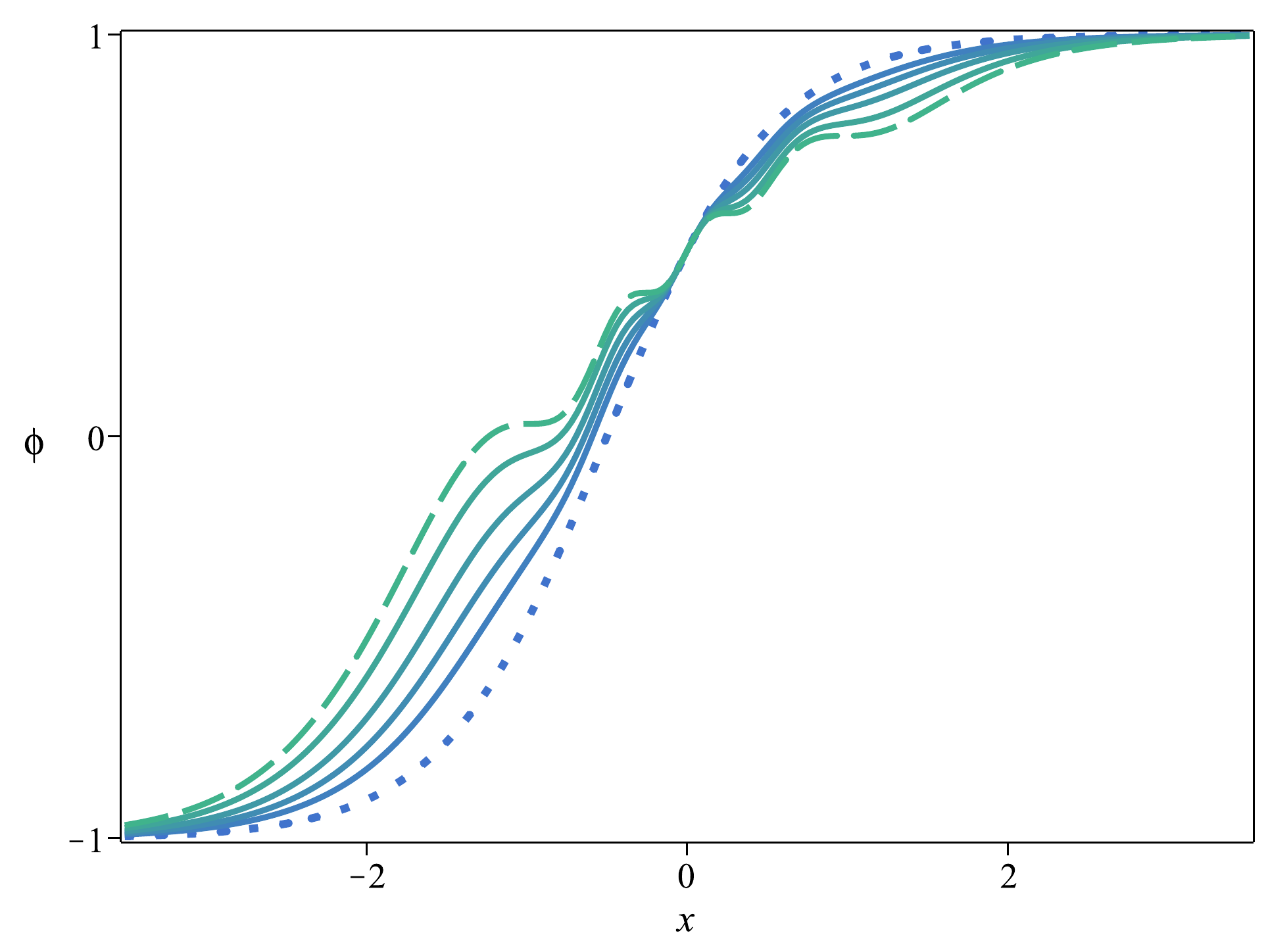}
		\includegraphics[width=4.2cm]{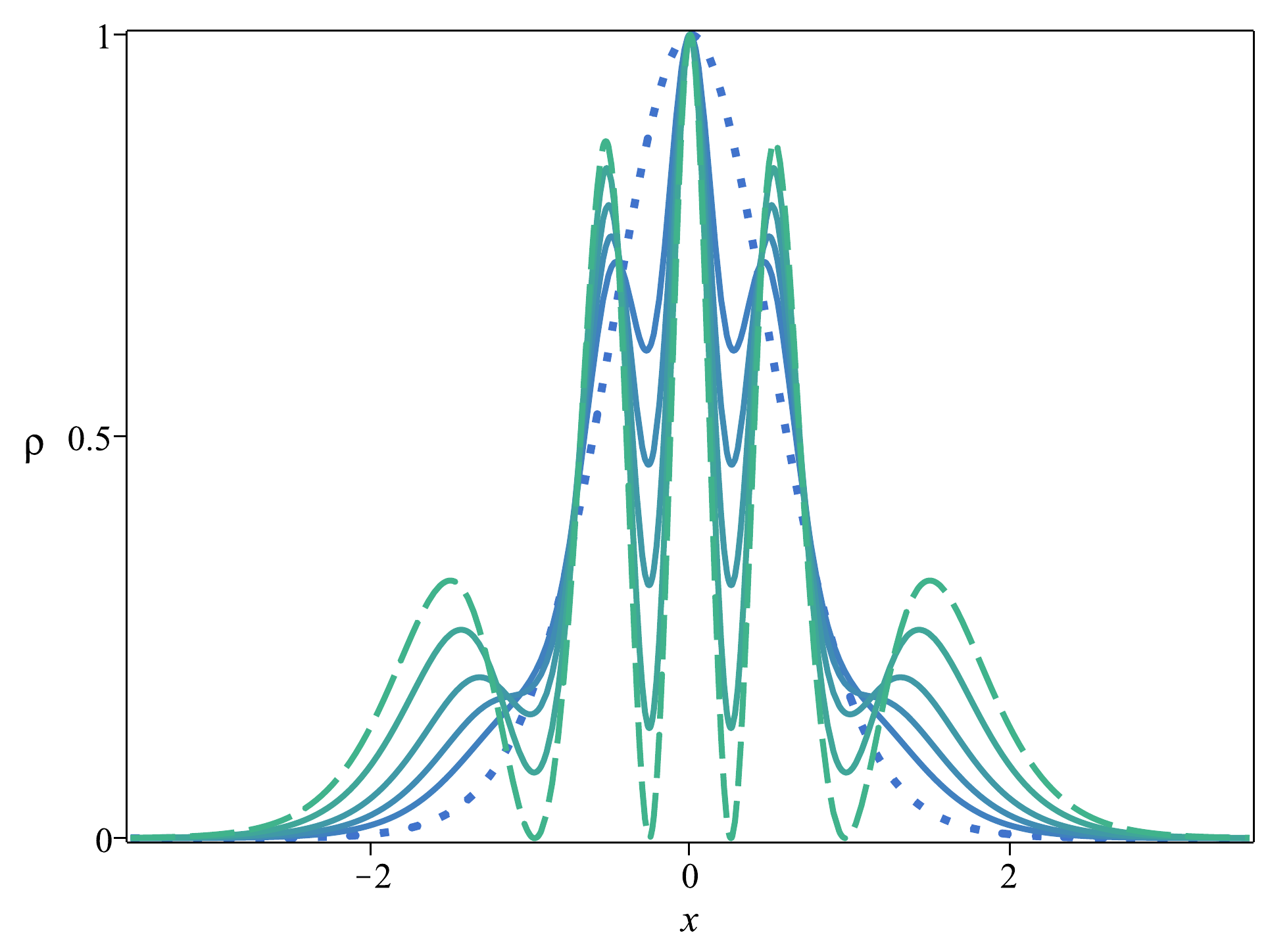}
		\includegraphics[width=4.2cm]{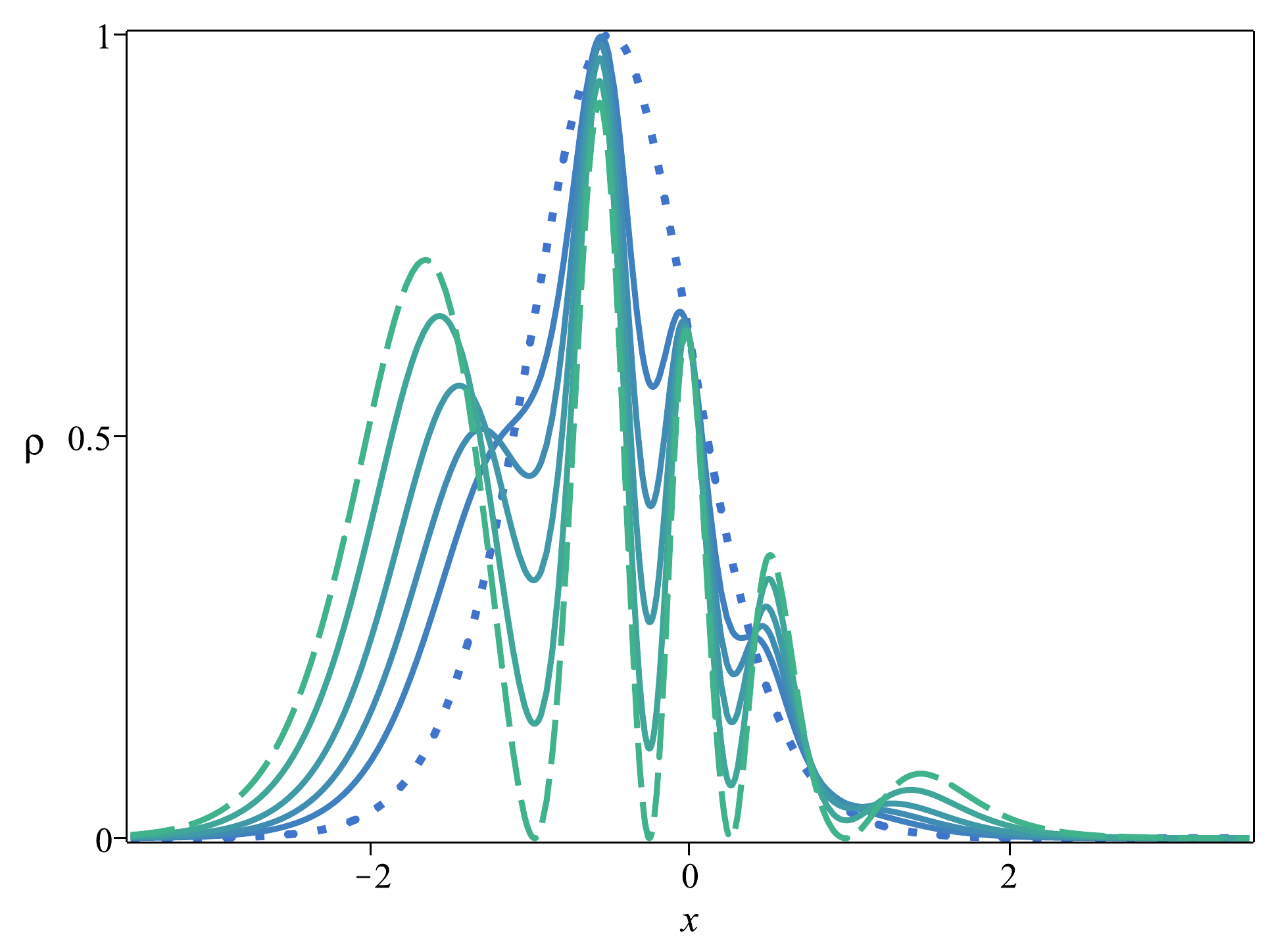}
		\caption{The solution $\phi$ (top) and the energy density $\rho_1$ (bottom) in Eq.~\eqref{solphi} for the geometrical coordinate in Eq.~\eqref{xi2} for several values of $\lambda$, with $\alpha=1$, $x_0=0$, $n=2$ and $\xi_0=0$ (left) and $1/2$ (right). The dotted lines represent the case with $\lambda=0$ and the dashed ones, the limit $\lambda\to\infty$.}
		\label{figsolrho2}
		\end{figure}

As we can see, the function in Eq.~\eqref{flambda1} induces a geometrical constriction in a single region of the kink, which we chose to be its center. We can modify this possibility by taking
\be\label{flambda2}
f(\psi)=\frac{1+\lambda}{1+\lambda\cos^2(n\pi\psi)},
\ee
where $n$ is an integer parameter. The periodicity of the cosine generalizes the previous possibility and induces other modifications, controlled by the integer $n$. To see this, one notices that the geometrical coordinate in Eq.~\eqref{xi} now takes the form
\be\label{xi2}
\begin{aligned}
	\xi &=\xi_0 + \frac{2+\lambda}{2(1+\lambda)}(x-x_0) \\
		&\hspace{4mm}+\frac{\lambda}{4\alpha(1+\lambda)}\Big(\textrm{Ci}(\zeta_n^+(x)) -\textrm{Ci}(\zeta_n^-(x))\Big),
\end{aligned}
\ee
where $\textrm{Ci}$ is the cosine integral function, and also, $\zeta_n^\pm(x) =2n\pi(1\pm\tanh(\alpha (x-x_0)))$. In this case, we get the profiles shown in Fig.~\ref{figsolrho2}. The internal structure induced by the function \eqref{flambda2} is richer than before, leading to $2n$ inflection points with vanishing derivative in the solution and $2n$ holes in the energy density. Notice that $\lambda$ makes a smooth transition from the standard configuration ($\lambda=0$) to the geometrically constrained one ($\lambda\to\infty$). As before, $\xi_0$ introduces an asymmetry in the system, shifting the inflection point in the solution and breaking the reflection symmetry ($x\to-x$) of the energy density, modifying the height of its peaks for $\lambda>0$.

One may also investigate other models, such as the sine-Gordon one, which arises for $g(\phi)=\sin\phi$. As it is known, it supports the family of solutions given by $\phi = \arcsin(\tanh(\xi)) + k\pi$, where $k\in\mathbb{Z}$, with energy density $\rho_1 = \sech^2(\xi)/f(\tanh(\alpha(x-x0)))$ and energy $E_1=2$. The modifications induced by the functions \eqref{flambda1} and \eqref{flambda2} in the sine-Gordon model are quite similar to the ones shown in Fig.~\ref{figsolrho1} and \ref{figsolrho2}, respectively, so we do not display them here. We can consider other periodic functions, instead of the cosine used in Eq. \eqref{flambda2}; this would lead to other results, but in a scenario similar to the above one, so we omit it here too. 

\section{Three fields} The previous model shows how to smoothly build internal structure and introduce asymmetry in kinklike solutions. As it is known, kinks can be used in the modeling of N\'eel wall. So, our procedure allows us to simulate geometrical constrictions in walls of the N\'eel type. One may also enlarge the previous model and introduce a third scalar field, $\chi$, with the Lagrangian density
\begin{equation}\label{l2}
	\begin{aligned}
		\LL &= \frac{1}{2}f(\psi)\pu \phi \partial^{\mu} \phi + \frac{1}{2}f(\psi)\pu \chi \partial^{\mu} \chi \\
			&\quad + \frac{1}{2}\pu \psi \partial^{\mu} \psi  - V(\phi,\chi,\psi).
	\end{aligned}
	\end{equation}
For constant and uniform $\psi$, this model becomes a two-field one, similarly to the to the cases considered in Refs.~\cite{bnrt1,bnrt2,bnrt3,bnrt4,bnrt5}. In this model, $\phi$ and $\chi$ may be used to model the two degrees of freedom of the so-called Bloch walls. As we shall see, the presence of the function $f$ allows to simulate geometrical constrictions in this type of domain walls. One can follow the lines of the previous model and find that the minimum energy configurations arise if the fields are static and the potential is written as
	\begin{equation}\label{potgen}
		V(\phi,\chi,\psi)=\frac12\frac{g_{\phi}^2}{f(\psi)} + \frac{1}{2}\frac{g_{\chi}^2}{f(\psi)} + \frac12 h_{\psi}^2,
	\end{equation}
where $g=g(\phi,\chi)$ and $h=h(\psi)$. In this case, if the first order equations
\bes\label{fo2}
\bal \label{fophi}
	&\frac{d\phi}{dx}=\pm \frac{g_{\phi}(\phi,\chi)}{f(\psi(x))},\\ \label{fochi}
	&\frac{d\chi}{dx}=\pm \frac{g_{\chi}(\phi,\chi)}{f(\psi(x))},\\ \label{fopsi}
	&\frac{d\psi}{dx}=\pm h_{\psi}(\psi(x)).
\eal
\ees
are satisfied, the energy of the solutions is minimized to $E=\left|W(\phi_\infty,\chi_\infty,\psi_\infty)-W(\phi_{-\infty},\chi_{-\infty},\psi_{-\infty})\right|$. Notice that Eq.~\eqref{fopsi} is equal to Eq.~\eqref{fo}. By using the geometrical coordinate \eqref{xi}, Eqs.~\eqref{fophi} and \eqref{fochi} can be rewritten as
\bes\label{foxi2}
\bal
	&\frac{d\phi}{d\xi}=\pm g_{\phi}(\phi,\chi),
	&\frac{d\chi}{d\xi}=\pm g_{\chi}(\phi,\chi).
\eal
\ees
Notice that $g$ depends on $\phi$ and $\chi$, so the above equations are coupled, with the coordinate $\xi$ depending on the solution $\psi(x)$ as in Eq.~\eqref{xi}. If the first order equations are valid, one can calculate the energy density with the expression $\rho=\rho_1+\rho_2$, where
\be\label{rhodw}
\rho_1 = \frac{g_\phi^2+g_\chi^2}{f(\psi)},\qquad \rho_2 = h_\psi^2.
\ee
To feed the function $f(\psi)$, we consider that $\psi$ is driven by the function in Eq.~\eqref{hpsi}, which leads to the solution and energy density in Eq.~\eqref{solpsi}. To investigate the Bloch wall, formed by $\phi$ and $\chi$, we take
\be
	g(\phi,\chi)=\phi - \frac{1}{3}\phi^3 - r\phi\chi^2.
\ee
Notice that the above expression involves a parameter $r$, which is non negative and controls the strength of the interaction between $\phi$ and $\chi$. In this case, the first order equations \eqref{foxi2} take the form
\bes\label{fobnrt}
\bal
	&\frac{d\phi}{d\xi}= 1-\phi^2-r\chi^2,\\
	&\frac{d\chi}{d\xi}= -2r\phi\chi.
\eal
\ees
One can use the above equations to show that this model supports the elliptic orbit (see Ref.~\cite{bflrorbit})
\be\label{orbit}
    \phi^{2}+\frac{r}{1-2r}\chi^{2}=1,
\ee
with $r\in (0,1/2)$, for the topological sector defined by the minima $v_{1}=(-1,0)$ and $v_{2}=(1,0)$, for the pair of fields $(\phi,\chi)$. This equation allows us to decouple Eqs.~\eqref{fobnrt} and find the analytical solutions
\be\label{solbnrt}
\phi(\xi) = \tanh(2r\xi)\quad\text{and}\quad \chi(\xi) = \sqrt{\frac{1-2r}{r}}\;\sech(2r\xi),
\ee
where $\xi$ is the geometrical coordinate defined by Eq.~\eqref{xi}, which depends on the specific form of $f(\psi)$, with $\psi(x)$ and its respective energy density $\rho_1(x)$ as in Eq.~\eqref{solpsi}. So, the Bloch wall, which is modulated by $\phi$ and $\chi$ is now in a geometry governed by $\xi$. The contribution of $\phi$ and $\chi$ in the energy density is given by Eq.~\eqref{rhodw}, which leads us to
\be\label{densbnrt}
\rho_{1}=\frac{4r}{f(\psi(x))}\,\sech^2(2r\xi)\Big(1-2r+\left(3r-1\right)\sech^2(2r\xi)\Big).
\ee
By integrating it, one gets $E_1=4/3$.

		\begin{figure}[t!]
		\centering
		\includegraphics[width=4.2cm]{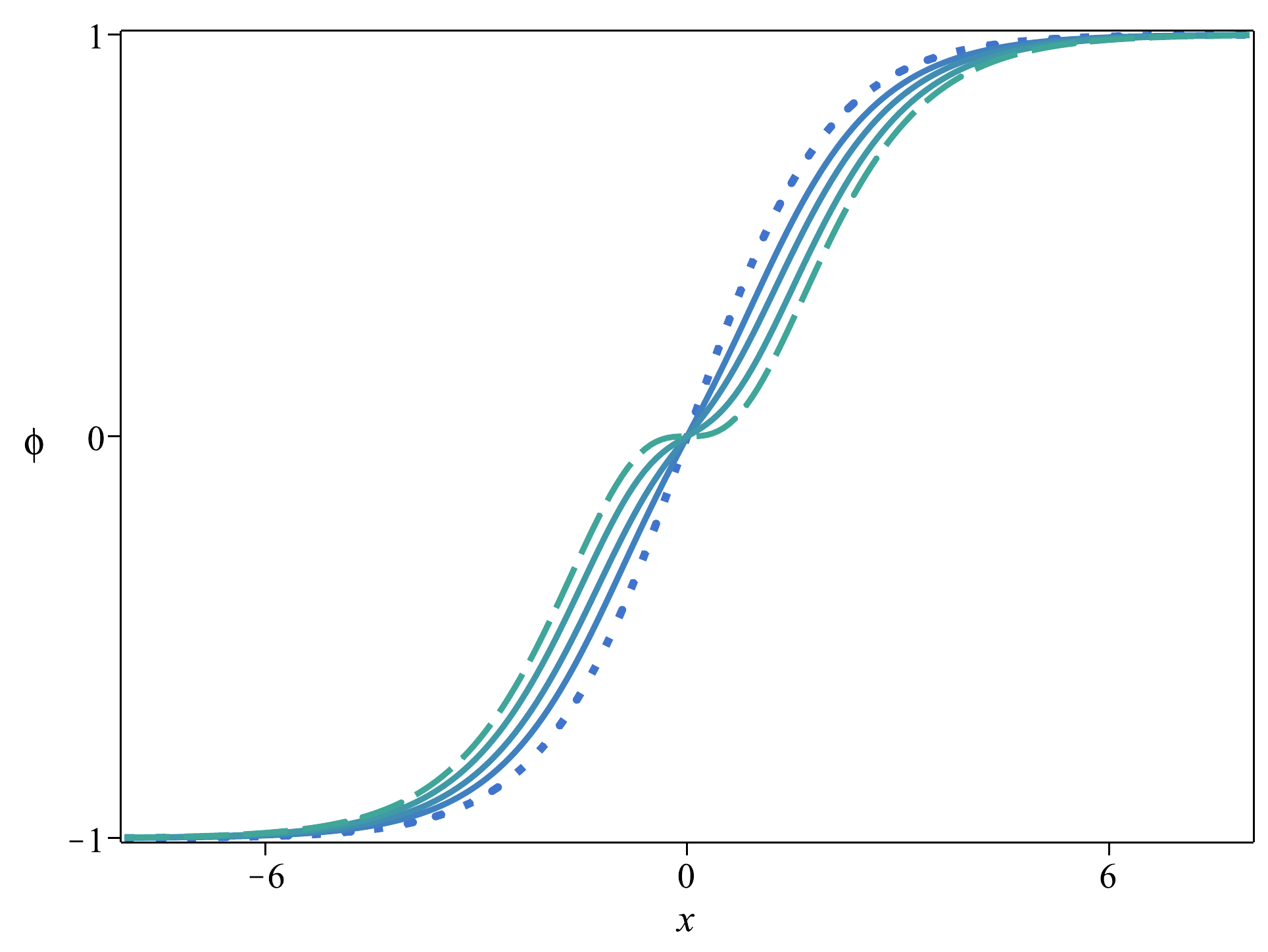}
		\includegraphics[width=4.2cm]{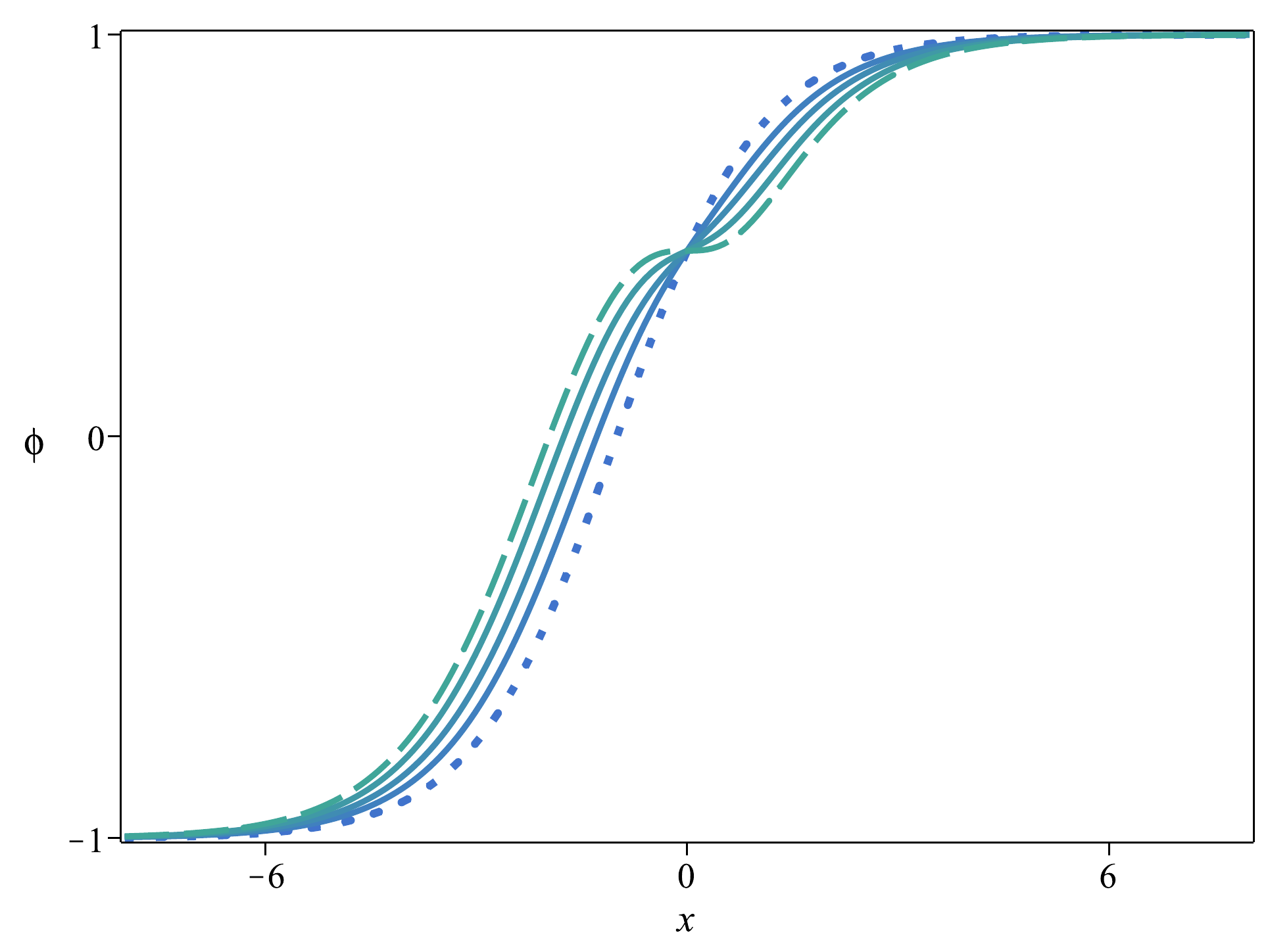}
		\includegraphics[width=4.2cm]{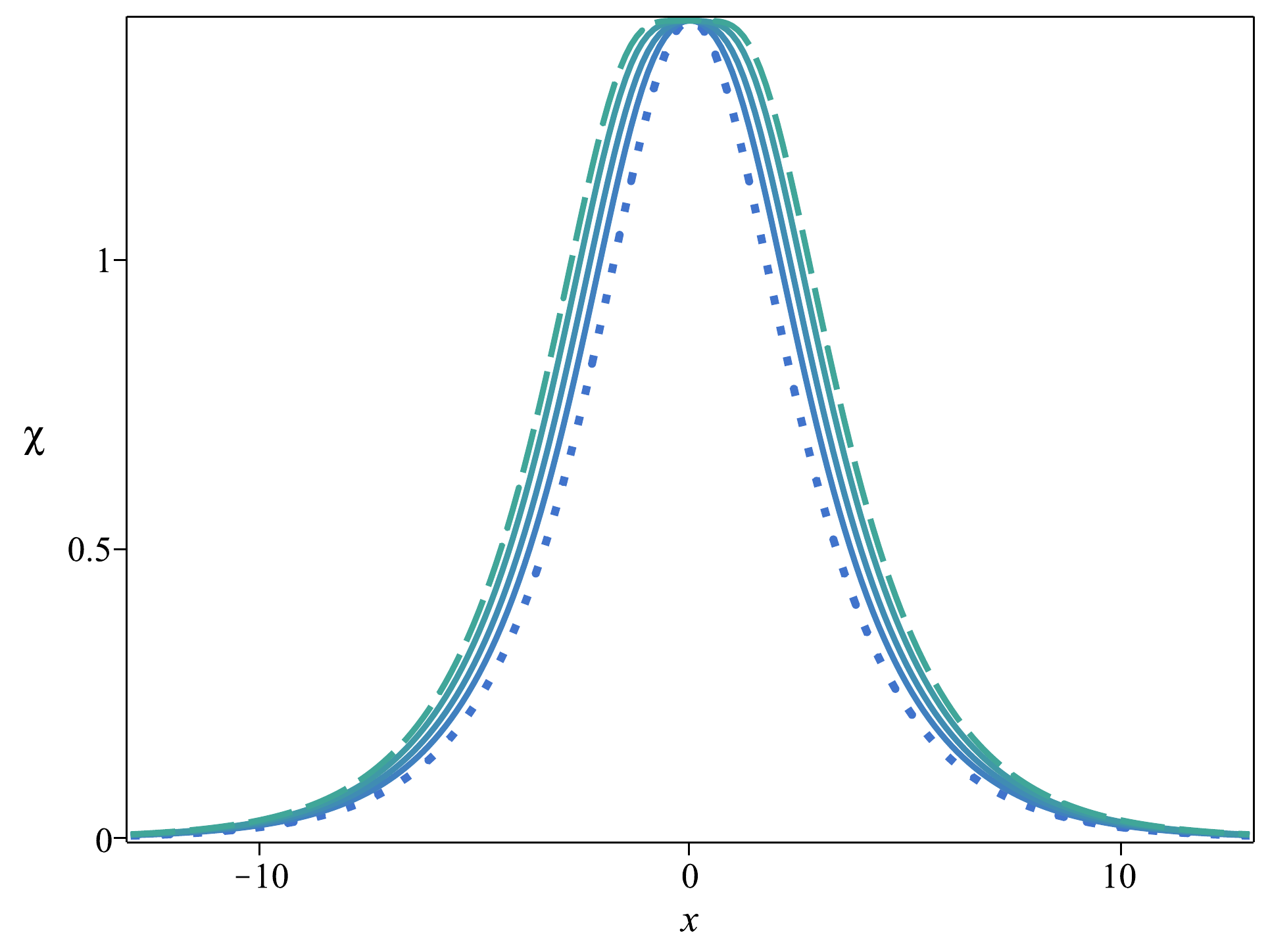}
		\includegraphics[width=4.2cm]{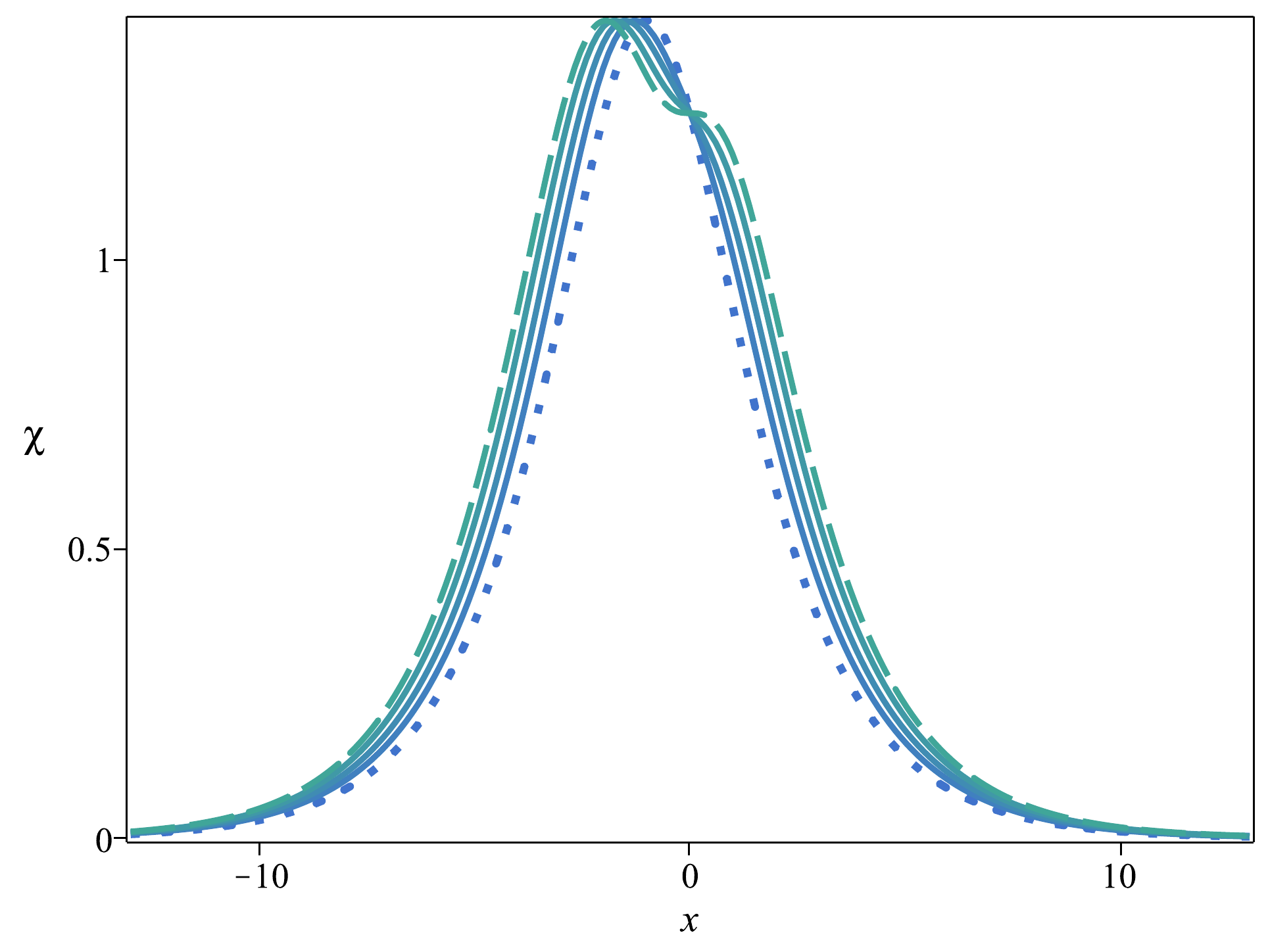}
		\includegraphics[width=4.2cm]{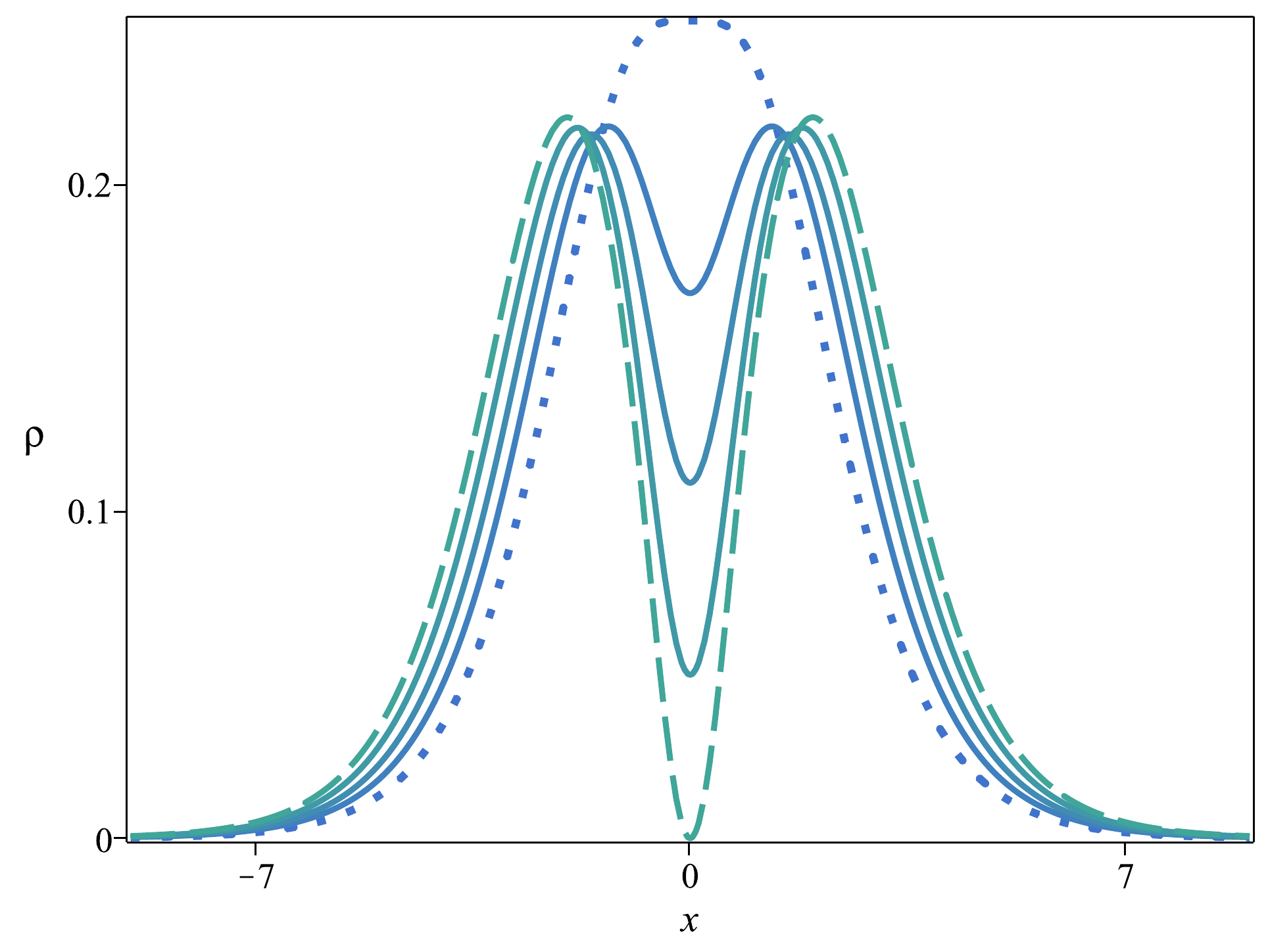}
		\includegraphics[width=4.2cm]{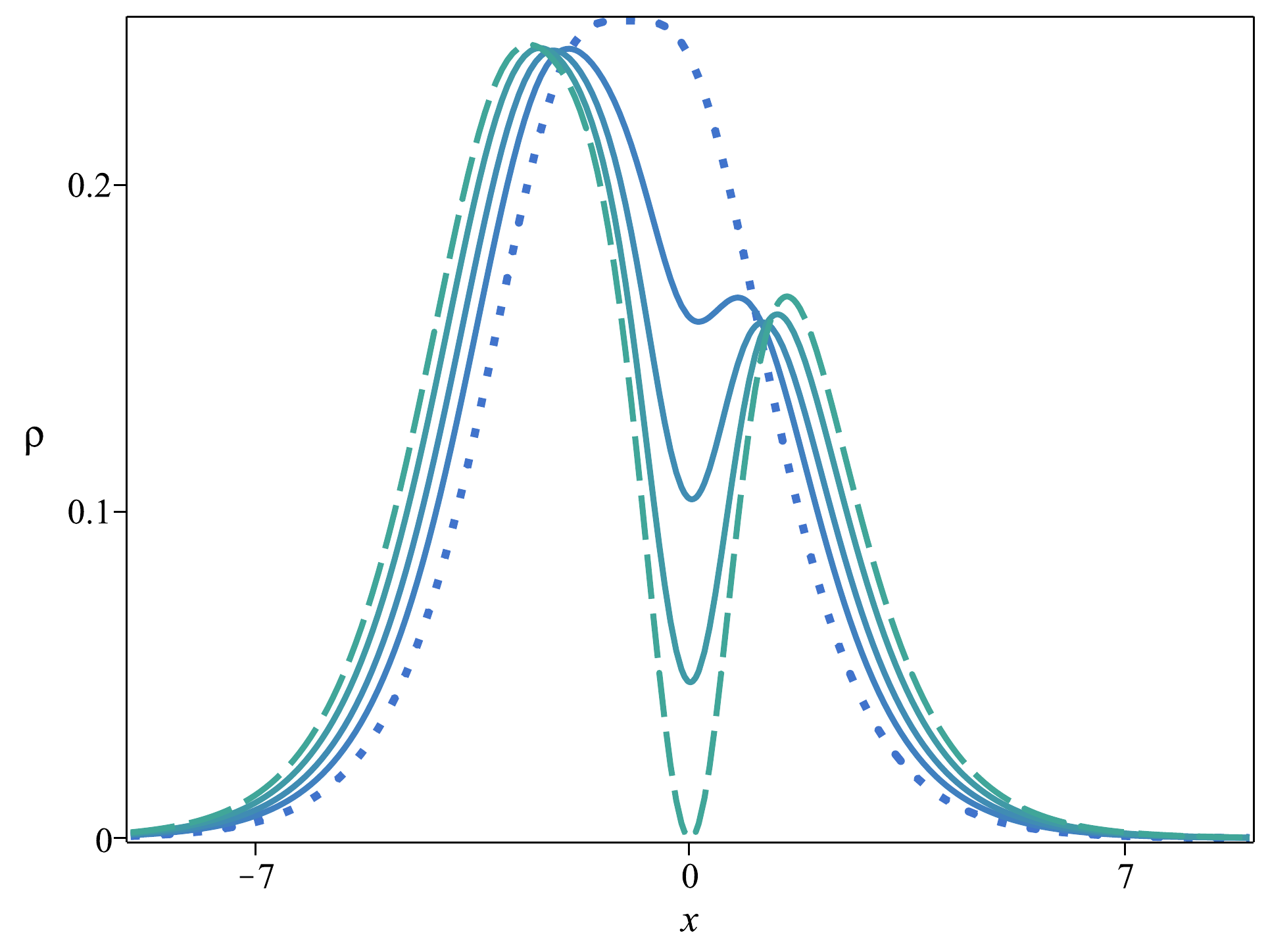}
		\caption{The solution $\phi$ (top), $\chi$ (middle) in Eq.~\eqref{solbnrt} and the energy density $\rho_1$ (bottom) in Eq.~\eqref{densbnrt} for the geometrical coordinate in Eq.~\eqref{xi1} for several values of $\lambda$, with $\alpha=1$, $x_0=0$, $r=1/4$ and $\xi_0=0$ (left) and $1/2$ (right). The dotted lines represent the case with $\lambda=0$ and the dashed ones do it for the limit $\lambda\to\infty$.}
		\label{figsolrho3}
		\end{figure}

To illustrate our procedure, we first consider the function $f(\psi)$ in Eq.~\eqref{flambda1}, for which the coordinate $\xi$ is given as in Eq.~\eqref{xi1}. In Fig.~\ref{figsolrho3}, we display the solutions $\phi$ and $\chi$ and their energy density $\rho_1$. Notice that, for $\xi_0=0$, in $\phi$, as $\lambda$ increases, the center of the kink tends to get an inflection point with vanishing derivative. The effect of the geometrical constriction simulated by the increasing of $\lambda$ in $\chi$ is different, making the central plateau become wider. In the energy density, we see that $\lambda$ controls the depth of the hole at its center. For $\xi\neq0$, we see that an asymmetry arises in the system. Here, we have a new effect: the lumplike solution $\chi$, becomes asymmetric. 

\begin{figure}[t!]
		\centering
		\includegraphics[width=4.2cm]{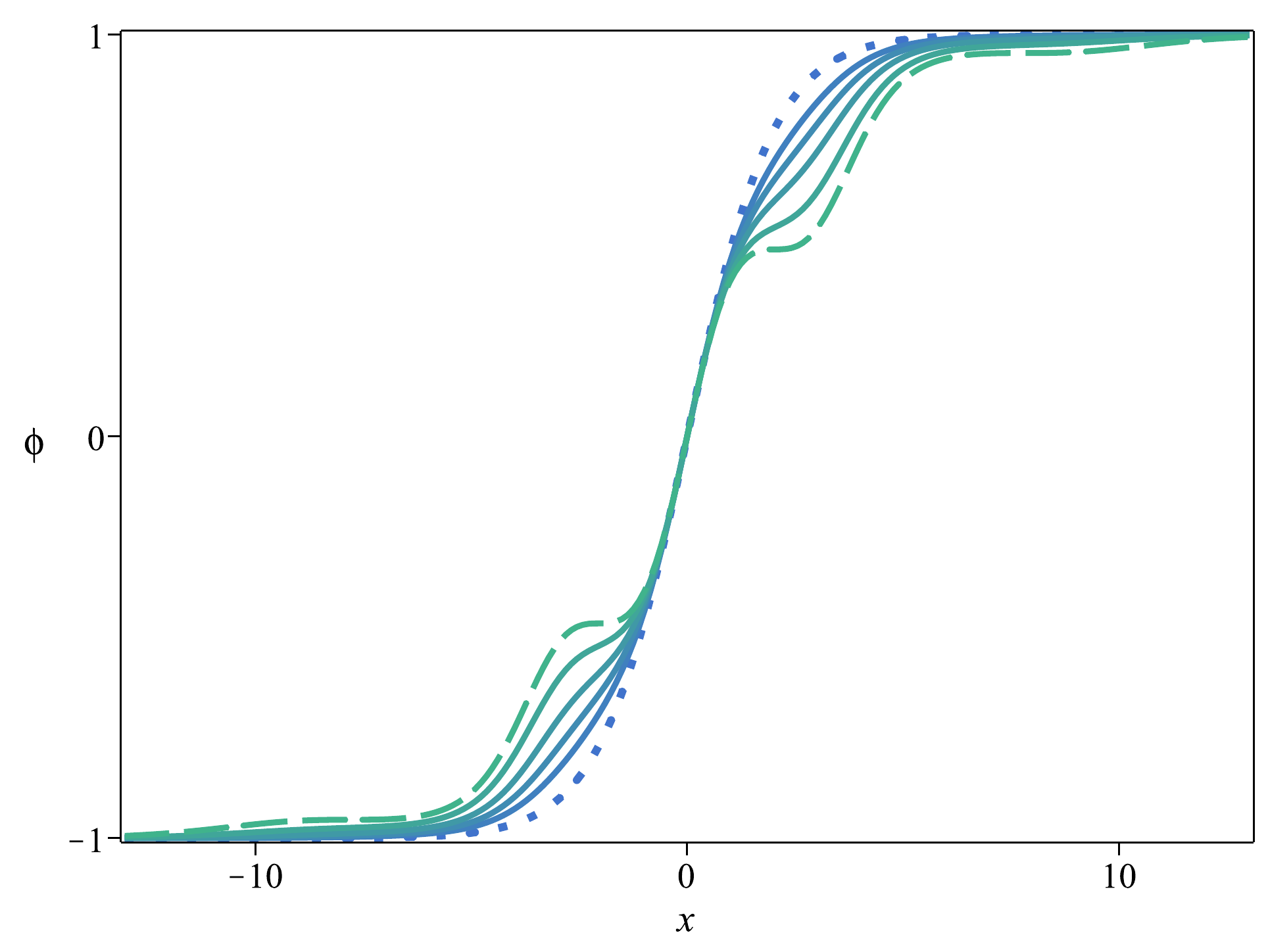}
		\includegraphics[width=4.2cm]{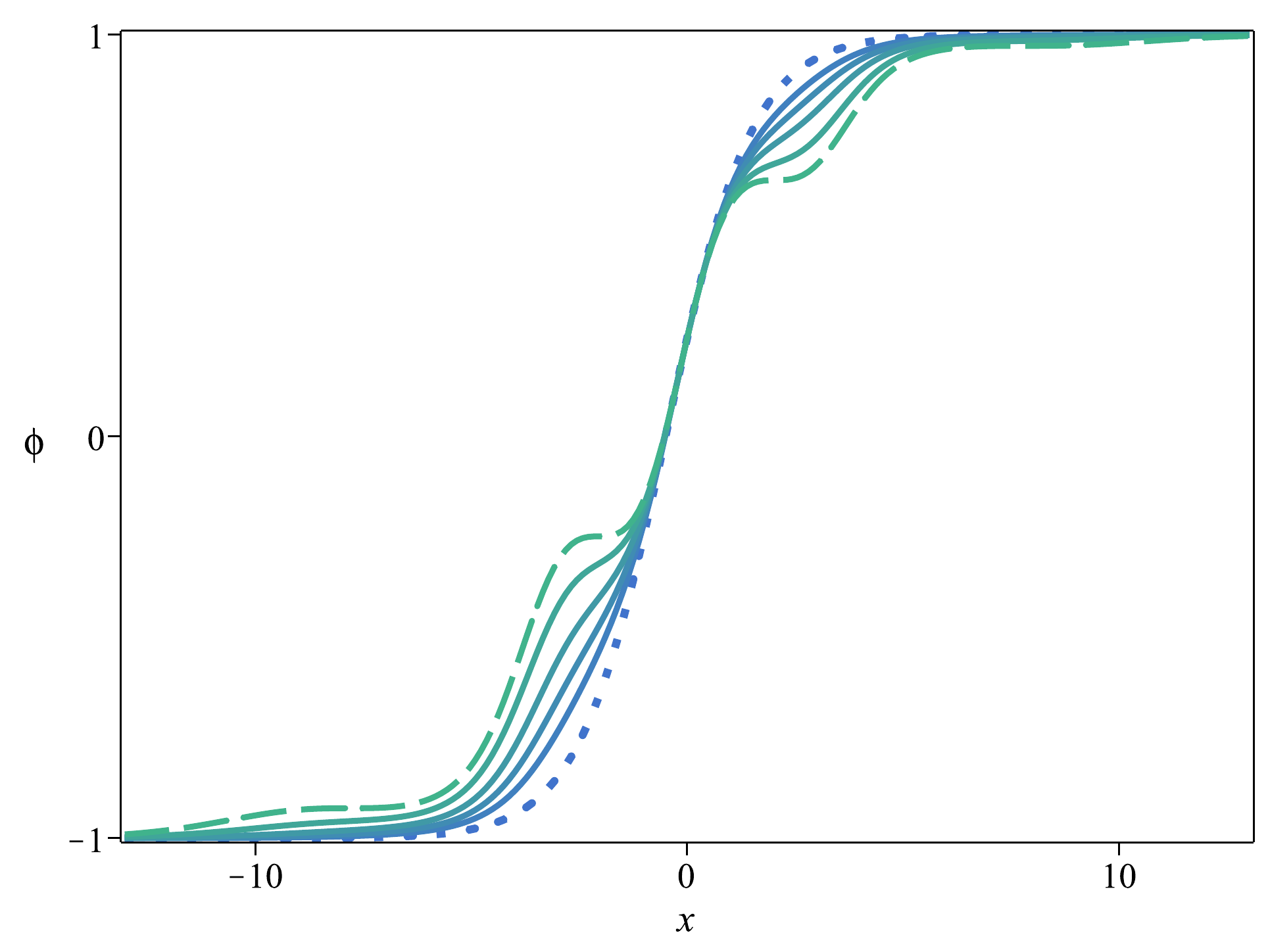}
		\includegraphics[width=4.2cm]{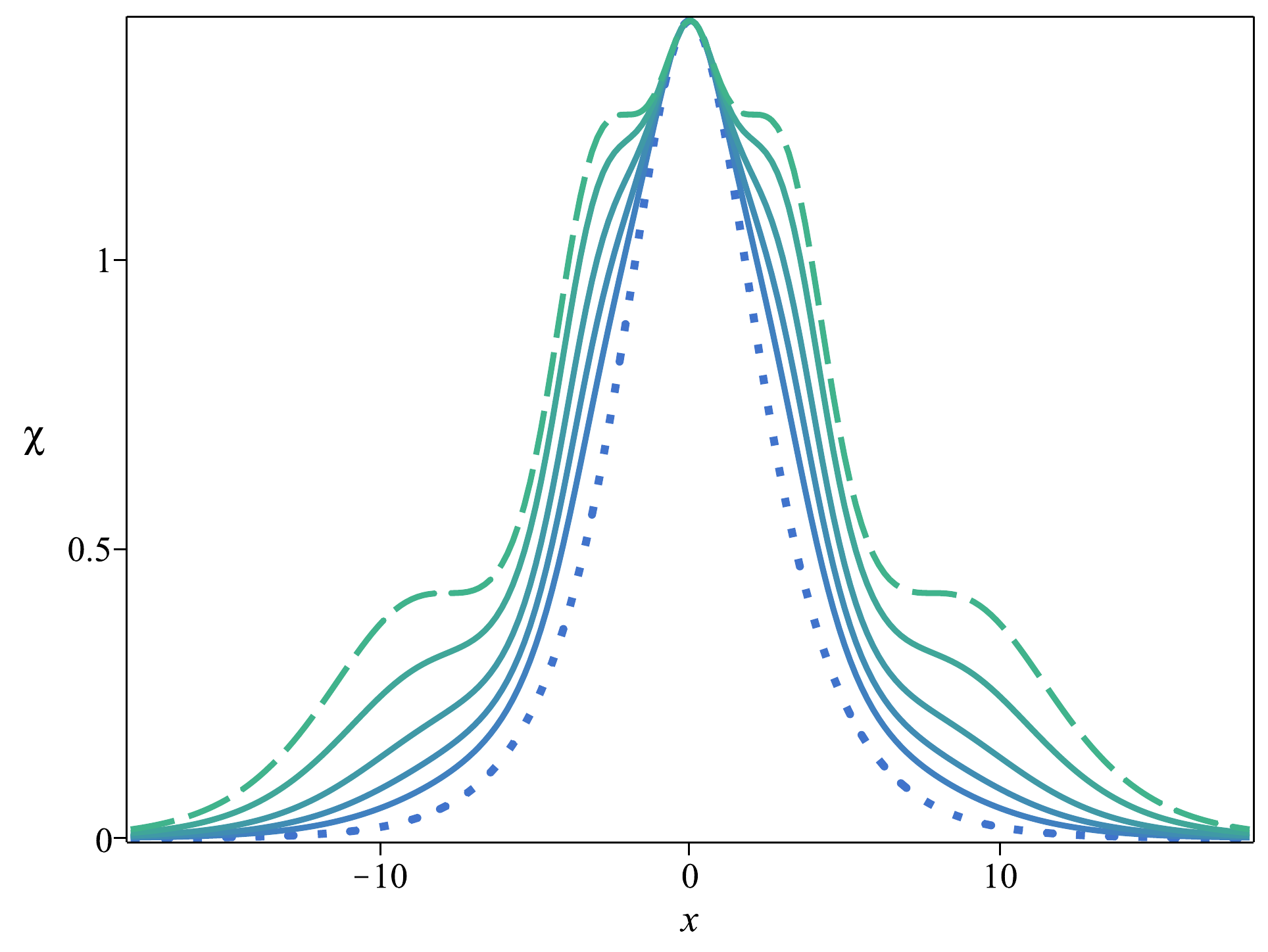}
		\includegraphics[width=4.2cm]{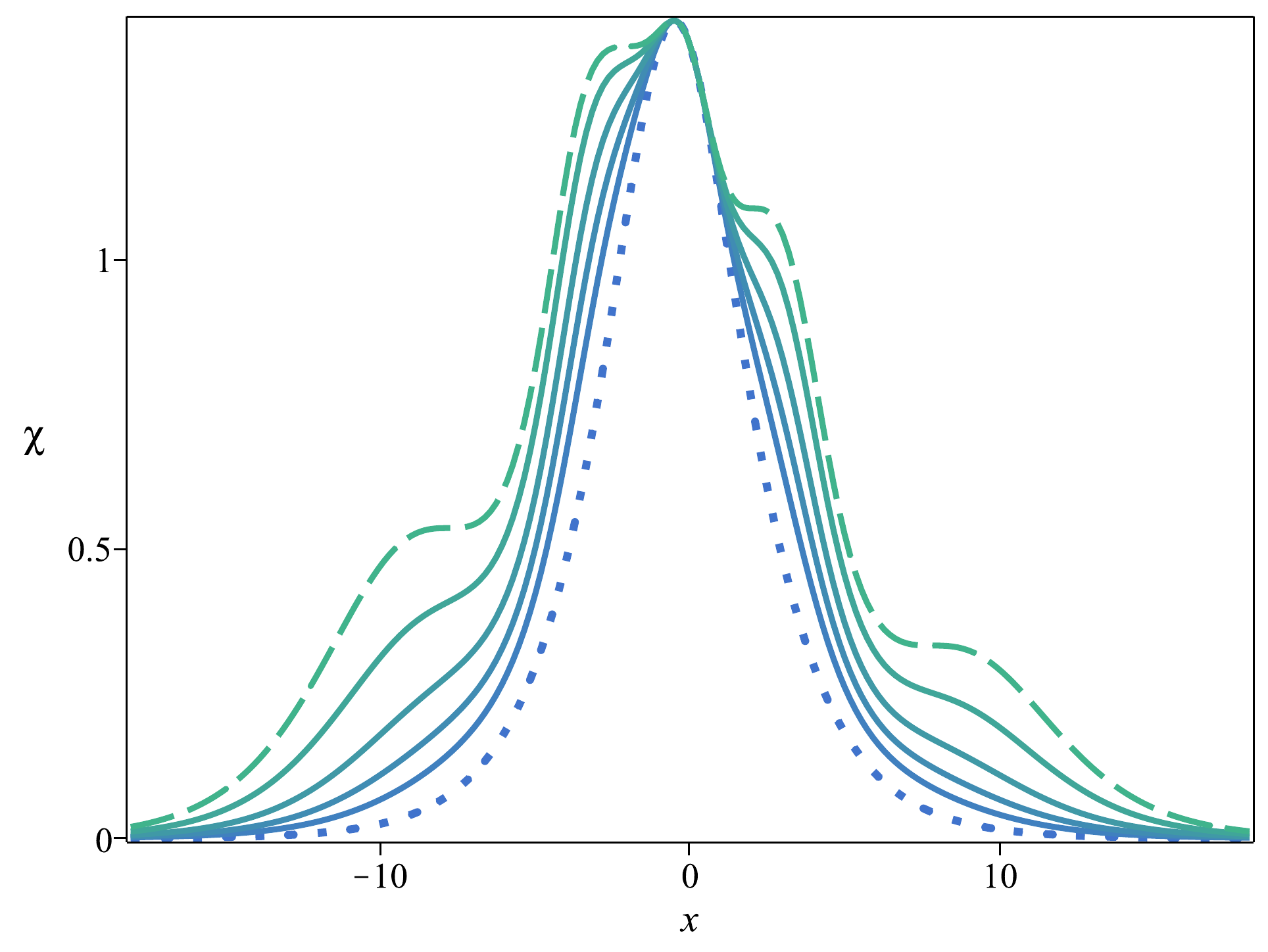}
		\includegraphics[width=4.2cm]{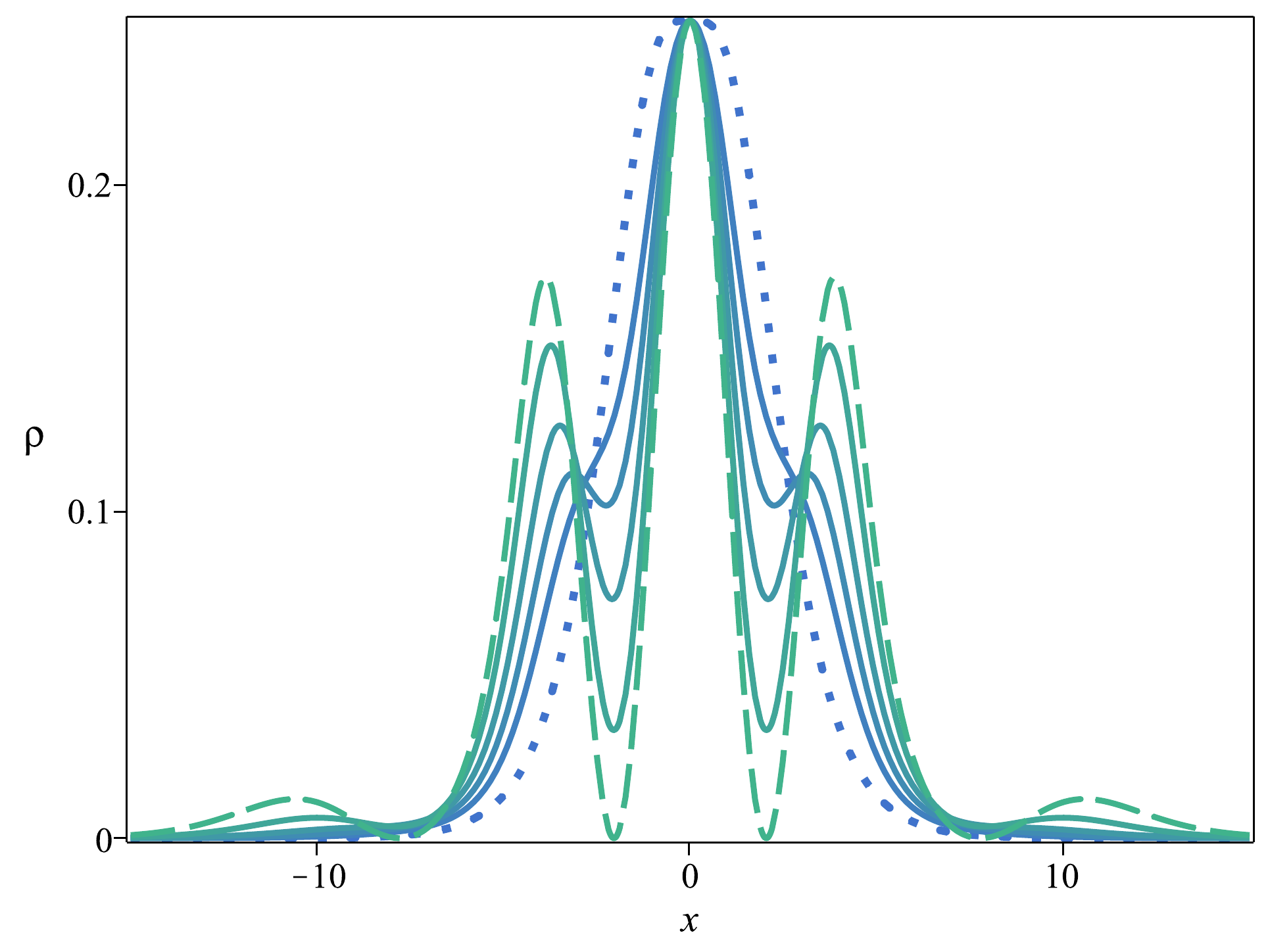}
		\includegraphics[width=4.2cm]{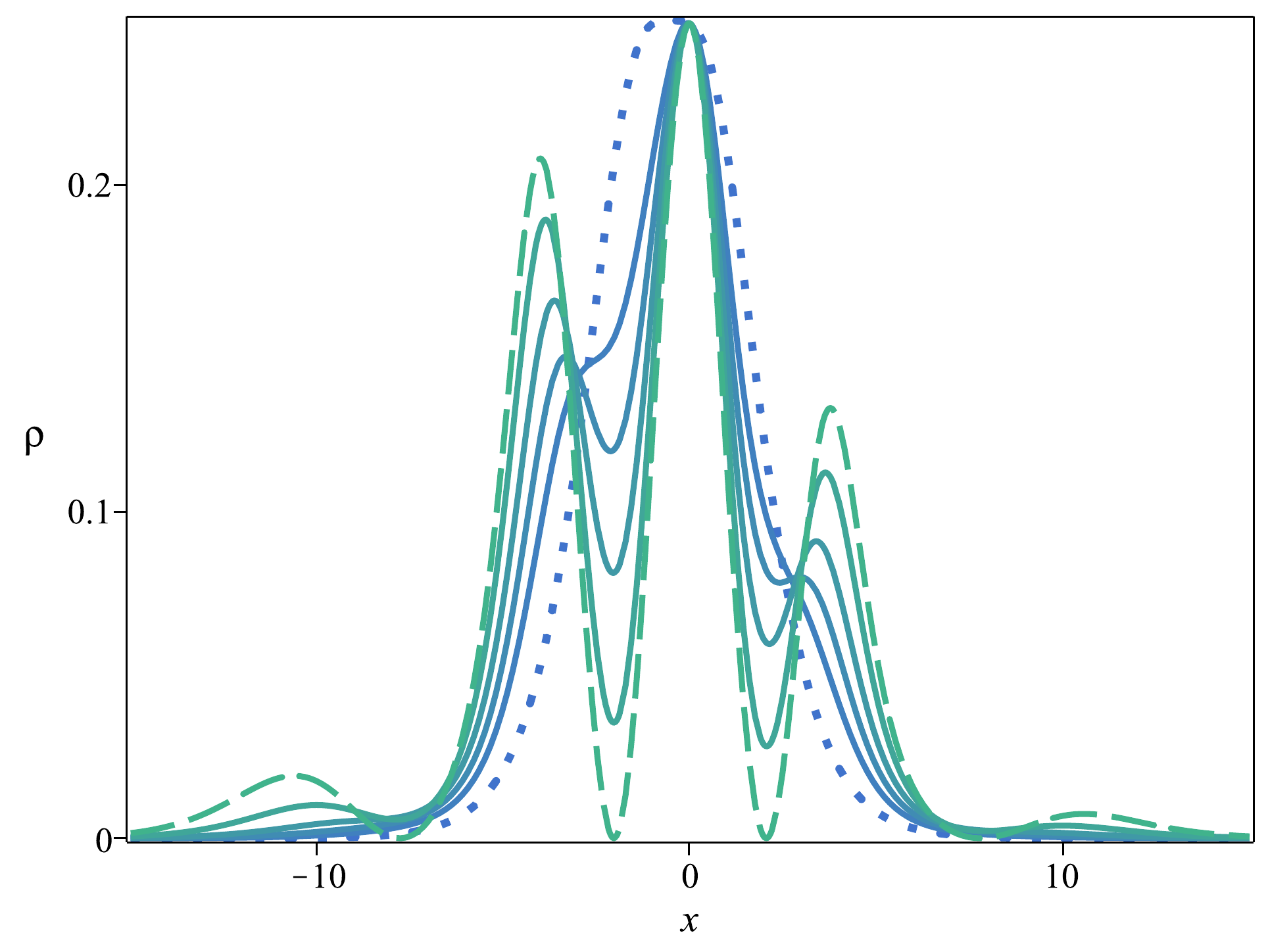}
		\caption{The solution $\phi$ (top), $\chi$ (middle) in Eq.~\eqref{solbnrt} and the energy density $\rho_1$ (bottom) in Eq.~\eqref{densbnrt} for the geometrical coordinate in Eq.~\eqref{xi2} for several values of $\lambda$, with $\alpha=1/8$, $x_0=0$, $r=1/4$, $n=2$ and $\xi_0=0$ (left) and $1/2$ (right). The dotted lines represent the case with $\lambda=0$ and the dashed ones do it for the limit $\lambda\to\infty$.}
		\label{figsolrho4}
		\end{figure}

To get a richer internal structure, we take $f(\psi)$ as in Eq.~\eqref{flambda2}, for which $\xi$ has the form in Eq.~\eqref{xi2}. The solutions $\phi$ and $\chi$, and their energy density $\rho_1$ can be seen in Fig.~\ref{figsolrho4}. As before, we have $2n$ points in the solutions and energy density that undergo geometrical constrictions as $\lambda$ gets larger. Notice that the asymmetric profile due to non null $\xi_0$ in the lumplike structure $\chi$ becomes richer, with several inflection points with vanishing derivative.

\section{Ending comments} In this work, we have shown that the inclusion of an extra scalar field with models in the class of Eqs.~\eqref{lagr} and \eqref{l2} in the study of kinks and domain walls can be used to smoothly build an internal structure and to introduce an asymmetry in the system. We have investigated the minimum energy configurations that are governed by first order equations. In this regime, the extra scalar field is independent and can be used as a source of a geometrical coordinate $\xi$, which has the form \eqref{xi}, and may simulate the effects of geometrical constrictions through the parameter $\lambda$ in the solutions. The results show that, as one increases $\lambda$, kinklike solutions get inflection points and lumplike solutions may get a wider plateaux at its center and/or inflection points, depending on the specific model.

The models here investigated also engender a second parameter, $\xi_0$, which is related to the presence of asymmetry in the solutions. For $\xi_0=0$, one has symmetric solutions around $x=x_0$. On the other hand, for $\xi_0\neq0$ and $\lambda>0$, one has asymmetric solutions and the energy density under the action of the geometrical constrictions gets shifted from the center of the structure. This feature gives rise to asymmetric localized structures of the lump type, which were not reported before in high energy physics.

Among the several issues related to the present study, an interesting possibility concerns applications to ferromagnetic materials. As already informed, recent investigation \cite{A3} have reported on the finding of chiral symmetry breaking between Bloch wall components due to an interlayer Dzyaloshinkii-Moriya interaction, and we think it is of current interest to further explore this path in the context of relativistic scalar fields. In Ref. \cite{BWB}, in particular, the study describes interesting way to adjust effects attributed to domain wall chirality induced by the Dzyaloshinskii–Moriya interaction at interfaces. Another possibility concerns the use of the above models for applications in thick brane scenarios, in particular to generalize the mechanism recently described in \cite{BW4} to control the internal structure of the corresponding warp factor and energy density. Another route of current interest concerns the presence of topological kinks in monolayer graphene, bilayer graphene and in other graphene-like classical wave systems \cite{KG}; see also the study of kink scattering in bilayer graphene in the presence of constriction \cite{BS}. We can also think of the inclusion of a second spatial dimension, to work with planar systems, searching for vortices and/or skyrmions, in particular, the Bloch-type skyrmions \cite{Sky} and the axially symmetric magnetic skyrmions of large size \cite{LS}. These and other related issues are currently under consideration, and we hope to report on some of them in the near future.

\vspace{1cm}

{\textit{ The present investigation is theoretical and the manuscript has no data to be associated with.}}

\acknowledgements{This work was partially financed by Conselho Nacional de Desenvolvimento Cient\'\i fico e Tecnol\'ogico (CNPq grant No. 303469/2019-6), by Federal University of Para\'iba (UFPB/PROPESQ/PRPG, project code PII13363-2020) and by Para\'iba State Research Foundation (FAPESQ-PB grant No. 0015/2019).



\begin{thebibliography}{99}
\bibitem{B1} N. Manton and P. Sutcliffe, \textit{Topological solitons.} Cambridge University Press (2004).
\bibitem{B2}T. Vachaspati, \textit{Kinks and domain walls.} Cambridge University Press (2006).
\bibitem{B3}E.J. Weinberg, \textit{Classical solutions in quantum field theory.} Cambridge University Press (2012).
\bibitem{BW00}L. Randall and R. Sundrum, Phys. Rev. Lett. 83, 4690 (1999).
\bibitem{BW1}O. DeWolfe, D.Z. Freedman, S.S. Gubser, and A. Karch, Phys. Rev. D 62, 046008 (2000).
\bibitem{BW2}C. Csaki, J. Erlich, T.J. Hollowood, and Y. Shirman, Nucl. Phys. B 581, 309 (2000).
\bibitem{BW3}V. Dzhunushaliev, V. Folomeev, and M. Minamitsuji, Rept. Prog. Phys. 73, 066901 (2010).
\bibitem{BW0}Yuan Zhong, JHEP 04, 118 (2021).
\bibitem{BW4}D. Bazeia and A.S. Lob\~ao Jr. Eur. Phys. J. C 82, L579 (2022).
\bibitem{BW5}Y. Zhong, K. Yang, and Y.-X. Liu, JHEP 09, 128 (2022).
\bibitem{BW6}Hao Geng, JHEP 2022, 24 (2022).
\bibitem{BW7}H. Geng, A. Karch, C. Perez-Pardavila, S. Raju, L. Randall, M. Riojas, and S. Shashi, JHEP  2022, 153 (2022).
\bibitem{KRB}A. Karch and L. Randall, JHEP 05, 008 (2001).
\bibitem{KCM1}P.-O. Jubert, R. Allenspach, and A. Bischof, Phys. Rev. B {69}, 220410(R) (2004).
\bibitem{A1}A. Vanhaverbeke, A. Bischof, and R. Allenspach, Phys. Rev. Lett. 101, 107202 (2008).
\bibitem{A2}T. Machon, G.P. Alexander, R.E. Goldstein, and A.I. Pesci, Phys. Rev. Lett. 117, 017801 (2016).
\bibitem{A3}S.D. Pollard, J.A. Garlow, K.-W. Kim, S. Cheng, K. Cai, Y. Zhu, and H. Yang, Phys. Rev. Lett. 125, 227203 (2020).
\bibitem{A4}J. Chen and S. Dong,
Phys. Rev. Lett. 126, 117603 (2021).
\bibitem{FE0}M. Hoffmann, F.P.G. Fengler, M. Herzig, T. Mittmann, B. Max, U. Schroeder, R. Negrea, P. Lucian, S. Slesazeck, and T. Mikolajick, Nature 565, 464 (2019).
\bibitem{MS1}M. Fiebig, T. Lottermoser, D. Meier, and M. Trassin, Nature Reviews Materials 1, 16046 (2016).
\bibitem{MS2}G.F. Nataf, M. Guennou, J.M. Gregg, D. Meier, J. Hlinka, E.K.H. Salje, and J. Kreisel, Nature Reviews Physics 2, 634 (2020). 
\bibitem{multikink} D.~Bazeia, M.A.~Liao and M.A.~Marques, 
Eur. Phys. J. Plus 135, 383 (2020).
\bibitem{BMM}D. Bazeia, A. Mohammadi, and D.C. Moreira, Phys. Rev. D 103, 025003 (2021).
\bibitem{bogo}E.B.~Bogomol'nyi, 
Sov. J. Nucl. Phys. 24, 449 (1976).
\bibitem{FE}M. Hoffmann, F.P.G. Fengler, M. Herzig, T. Mittmann, B. Max, U. Schroeder, R. Negrea, P. Lucian, S. Slesazeck, and T. Mikolajick, Nature  565, 464 (2019).
\bibitem{FM}W. Jiang P. Upadhyaya, W. Zhang, G. Yu, M. Benjamin Jungfleish, F.Y. Fradin, J.E. Pearson, Y. Tserkovnyak, K.L. Wang, O. Heinonen, S.G.E. Tevelthuis, and A. Hoffmann, Science 349, 283 (2015).
\bibitem{bnrt1}D.~Bazeia, M.J.~dos Santos and R.F.~Ribeiro, 
Phys. Lett. A 208, 84 (1995).
\bibitem{bnrt2}M.A. Shifman, M.B. Voloshin, Phys. Rev. D 57, 2590 (1998).
\bibitem{bnrt3}D. Bazeia and F.A. Brito, Phys. Rev. D 61, 105019 (2000).
\bibitem{bnrt4}A. Alonso-Izquierdo, M.A. Gonzalez Leon, J.M. Guilarte, Phys. Rev. D 65, 085012 (2002).
\bibitem{bnrt5}P.P. Avelino, D. Bazeia, R. Menezes, and J.C.R.E. Oliveira,  Phys. Rev. D 79, 085007 (2009).
\bibitem{bflrorbit}D.~Bazeia, W.~Freire, L.~Losano and R.F.~Ribeiro, 
Mod. Phys. Lett. A {\bf17}, 1945 (2002).
\bibitem{BWB}G. Chen, T. Ma, A.T. N’Diaye, H. Kwon, C. Won, Y. Wu, and A.K. Schmid, Nature Communications 4, 2671 (2013).
\bibitem{KG}Z. Wang, S. Cheng, X. Liu, and H. Jiang,
Nanotechnology 32, 40 (2021).
\bibitem{BS}N. Benchtaber, D. Sánchez, and L. Serra, Phys. Rev. B 104, 155303 (2021).
\bibitem{Sky}A.N. Bogdanov and C. Panagopoulos, 
Nature Reviews Physics 2, 492 (2020).
\bibitem{LS}S. Komineas, C. Melcher, and S. Venakides, Physica D 418, 132842 (2021).
\end{thebibliography}
\end{document}